\documentclass[12pt]{article}
\makeatletter \@addtoreset{equation}{section} \makeatother
\renewcommand{\theequation}{\thesection.\arabic{equation}}
\newcommand{\ba}{\begin{array}}
\newcommand{\ea}{\end{array}}
\newcommand{\beq}{\begin{equation}}
\newcommand{\eeq}{\end{equation}}
\newcommand{\bea}{\begin{eqnarray}}
\newcommand{\eea}{\end{eqnarray}}

\def\bce{\begin{center}}
\def\ece{\end{center}}

\def\nonu{\nonumber}

\def\pa{\partial}
\def\al{\alpha}
\def\be{\beta}

\def\de{\delta}

\def\ep{\epsilon}

\def\th{\theta}

\def\Kop{{\displaystyle \mathop{K}^{\circ}}{}}
\def\Gop{{\displaystyle \mathop{g}^{\circ}}{}}
\def\eop{{\displaystyle \mathop{e}^{\circ}}{}}

\def\eps6{{\displaystyle \mathop{\epsilon}^{6}}{}}
\def\g6{{\displaystyle \mathop{g}^{6}}{}}

\def\nab6{{\displaystyle \mathop{\nabla}^{6}}{}}

\def\0{{\sst{(0)}}}
\def\1{{\sst{(1)}}}
\def\2{{\sst{(2)}}}
\def\3{{\sst{(3)}}}
\def\4{{\sst{(4)}}}
\def\5{{\sst{(5)}}}
\def\6{{\sst{(6)}}}
\def\7{{\sst{(7)}}}
\def\8{{\sst{(8)}}}

\def\ba{\begin{array}}
\def\ea{\end{array}}
\def\beq{\begin{equation}}
\def\eeq{\end{equation}}
\def\be{\begin{equation}}
\def\ee{\end{equation}}

\def\eps{\epsilon}

\def\th{{\theta}}

\def\ba{\begin{array}}
\def\ea{\end{array}}
\def\beq{\begin{equation}}
\def\eeq{\end{equation}}
\def\be{\begin{equation}}
\def\ee{\end{equation}}

\def\eps{\epsilon}

\def\th{{\theta}}

\def\Kop{{\displaystyle \mathop{K}^{\circ}}{}}
\def\eop{{\displaystyle \mathop{e}^{\circ}}{}}
\def\Dop{{\displaystyle \mathop{D}^{\circ}}{}}
\def\Gop{{\displaystyle \mathop{g}^{\circ}}{}}
\def\eop{{\displaystyle \mathop{e}^{\circ}}{}}

\def\eps6{{\displaystyle \mathop{\epsilon}^{6}}{}}

\def\nab6{{\displaystyle \mathop{\nabla}^{6}}{}}

\newcommand{\bean}{\begin{eqnarray*}}
\newcommand{\eean}{\end{eqnarray*}}

\begin{document}
\thispagestyle{empty} \addtocounter{page}{-1}
   \begin{flushright}
PUPT-2386 \\
\end{flushright}

\vspace*{1.3cm}
  
\centerline{ \Large \bf   
Towards New Membrane Flow 
from de Wit-Nicolai Construction}
\vspace*{1.5cm}
\centerline{{\bf Changhyun Ahn  $^{\dagger}$ $^{\ast}$
\footnote{On leave from the Department of Physics, Kyungpook National University, Taegu
  702-701, Korea and 
address until Aug. 31, 2011:
Department of Physics, Princeton University, Jadwin Hall, 
Princeton, NJ 08544, USA},
Jinsub Paeng $^{\ast}$ 
 {\rm and} Kyungsung Woo $^{\ast}$ }
} 
\vspace*{1.0cm} 
\centerline{\it $^{\dagger}$ 
Department of Physics, Princeton University, Jadwin Hall, 
Princeton, NJ 08544, USA}
\centerline{\it  $^{\ast}$
Department of Physics, Kyungpook National University, Taegu
702-701, Korea} 
\vspace*{0.8cm} 
\centerline{\tt ahn@knu.ac.kr \qquad 
jdp2r@virginia.edu \qquad wooks@knu.ac.kr
} 
\vskip2cm

\centerline{\bf Abstract}
\vspace*{0.5cm}

The internal 4-form field strengths with 7-dimensional indices  
have been constructed by de Wit
and Nicolai in 1986. They are determined by 
the following six quantities: the 
56-bein of 4-dimensional ${\cal N}=8$ gauged supergravity, 
the Killing vectors on  the round seven-sphere, the covariant
derivative acting on these Killing vectors, the warp factor, 
the field strengths with
4-dimensional indices and the 7-dimensional metric. 

In this paper, 
by projecting out the remaining mixed 4-form field strengths 
in an $SU(8)$ tensor that appears in the variation of spin
$\frac{1}{2}$ fermionic sector,  we also write down them explicitly
in terms of some of the above quantities. 
For the known critical points, the ${\cal N}=8$ $SO(8)$ point and 
the nonsupersymmetric $SO(7)^{+}$  point,  
we reproduce the corresponding 11-dimensional uplifts by computing the
full nonlinear expressions.  
Moreover, we find  out 
the 11-dimensional lift of 
the nonsupersymmetric $SO(7)^{+}$ invariant flow.
We decode their implicit formula for the first time 
and the present work will provide how to
obtain the new supersymmetric or nonsupersymmetric membrane flows in 11-dimensions.   

\baselineskip=18pt
\newpage
\renewcommand{\theequation}
{\arabic{section}\mbox{.}\arabic{equation}}

\section{Introduction}

The truncation of 11-dimensional supergravity on the seven-sphere 
to the massless sector is equivalent to the 4-dimensional gauged
${\cal N}=8$ supergravity \cite{dN87}. This enables us to write down
the full nonlinear metric ansatz 
directly from the vacuum expectation values of the scalar
and pseudoscalar fields of 4-dimensional gauged ${\cal N}=8$
supergravity \cite{dWNW}, together with both warp factor and the Killing vectors on
the seven-sphere.
The 7-dimensional inverse metric is generated from the $SU(3)$-singlet
vacuum expectation values of 4-dimensional gauged ${\cal N}=8$
supergravity \cite{AI02}. 
Although this metric is written in terms of 
the rectangular coordinates, the standard metric in terms of
7-dimensional global coordinates is recovered. 
For the 4-form field strengths, some of the components of 
the full nonlinear ansatz are 
found by de Wit and Nicolai \cite{dN87} but the remaining ones 
of the 4-form components, where the four indices contain both
the internal 7-dimensional indices and 4-dimensional indices,
are not known so far.
There exist some previous works \cite{CLP99,CLP00} where 
the full nonlinear metric ansatz is used but the 4-form ansatz is not
used(the 4-forms are determined by brute force)
because there exist only partial informations on these 4-form field
strengths and it is difficult to decode their implicit formula 
for practical use.

In this paper, we reexamine the work of de Wit and Nicolai \cite{dN87} 
and would like to see whether the full nonlinear ansatz for 4-form
field strengths provide a master equation for the 11-dimensional solution.
The situation when we deal with the 4-forms is more complicated than 
the one with the
metric because the $SU(8)$ covariance for the theory requires the
five-fold product of the `generalized' vielbein $e^m_{ij}$ where $i,j$
are $SU(8)$ indices and $m$ is the 7-dimensional curved index. 
For the full nonlinear
metric ansatz, the two-fold product of them is needed. 
This is the reason why the explicit computations for the 4-forms
completely are not known so far during last 25 years. 
The supersymmetric flow solutions for $SU(3) \times U(1)_R$ invariant
flow \cite{CPW} and $G_2$ invariant flow \cite{AI} are found by taking
the appropriate 4-forms ansatz via the symmetry of the theory rather
than using the formula of \cite{dN87}. Of course, the full nonlinear
metric ansatz \cite{dWNW} are used here.  

For the two-fold product of generalized vielbeins in
the full nonlinear metric, 
it is nontrivial to write down the rectangular coordinates
in terms of the 7-dimensional curved coordinates or frame coordinates
but it is straightforward to express the 7-dimensional inverse metric
in terms of the rectangular coordinates, as a first step.
For the 4-forms, the data from 4-dimensional gauged ${\cal N}=8$
supergravity goes into the generalized vielbein. The five copies among
these generalized vielbeins with
an appropriate $SU(8)$ indices make the explicit computation complicated. 
As we multiply them successively, the expressions are getting more involved.
For the time being, we consider and focus on the simplest cases where the
4-dimensional data looks very simple: $SO(8)$ critical point,
$SO(7)^{+}$ critical point and $SO(7)^{+}$ invariant flow.
Note that there exist three basic representations of $SO(8)$, the
vector ${\bf 8}_v$,
the left handed spinor ${\bf 8}_{-}$ and the right handed spinor
${\bf 8}_{+}$. The embeddings of $SO(7)^{\pm}$ in $SO(8)$ for
representation ${\bf 8}_v$ are the same but those for the
representation ${\bf 8}_{\pm}$ are different from those for the
representation ${\bf 8}_{\mp}$. 
For the $SO(7)^{+}$ critical point, the scalar has nonzero vauum
expectation values while  for 
the $SO(7)^{-}$ critical point, the pseudo scalar has nonzero vauum
expectation values.

In section 2, we review the main results of 
the de Wit and Nicolai's construction, and 
obtain the mixed 4-form field strengths newly. 

In section 3, we apply the formula of section 2 to 
the two critical points and the nonsupersymmetric flow and 
find out the corresponding 11-dimensional solutions.
Some of them are previously known.

In section 4, we summarize the present work and 
comment on the future directions.

In the Appendices, we present the detailed expressions for three cases
in section 3 and describe the supersymmetry checking for the
11-dimensional $SO(7)^{+}$ flow solution.

\section{de Wit-Nicolai construction}

In this section, we describe the de Wit and Nicolai construction which 
provides the full nonlinear ansatz for the 4-form field
strengths. Later, we continue to apply 
their construction to the mixed 4-forms which is necessary for the
11-dimensional uplift of 4-dimensional domain wall solutions.  

\subsection{The 4-form field strengths}

In this section, we describe the relevant parts for the full nonlinear ansatz
\cite{dN87} for the 11-dimensional 4-form field 
strengths \cite{CJS} with internal 7-dimensional indices 
in terms of the data of 4-dimensional
gauged ${\cal N}=8$ supergravity \cite{dN,dN1}. 
For those who are interested in the
details of this construction, we refer to the original paper
\cite{dN87} by de
Wit and Nicolai.

The variation of spin-1 field of 11-dimensional supergravity 
contains the generalized vielbein that
has the coefficient function, which depends on only 4-dimensional space-time, 
in front of the Killing vector on the round seven-sphere.
It turns out, from ${\cal N}=8$ transformation rule for vector field,  
that this coefficient function  
can be written as the following four-dimensional quantity \cite{dN87}
\bea
w_{ij}^{\,\,IJ}(x) \equiv u_{ij}^{\,\,IJ}(x) + v_{ijIJ}(x),
\label{w}
\eea
where 
$ u_{ij}^{\,\,IJ}$ and $v_{ijIJ}$
fields are $28 \times 28$ matrices of ${\cal N}=8$ gauged supergravity
that depend on the 4-dimensional
curved space-time $x^{\mu}$.
Here the $SU(8)$ indices $[ij]$ are antisymmetrized 
and the $SO(8)$ indices $[IJ]$ are also antisymmetrized.
We will use these properties all the times.
All the indices run from $1$ to $8$.
The complex conjugation of $w_{ij}^{\,\,IJ}$  
can be obtained by raising or lowering  
the indices. 
That is, $(w_{ij}^{\,\,IJ})^{\ast} = w^{ij}_{\,\,IJ} =
(u_{ij}^{\,\,IJ})^{\ast} + (v_{ijIJ})^{\ast}
=u^{ij}_{\,\,IJ} + v^{ijIJ}$ from the 4-dimensional gauged ${\cal
  N}=8$ supergravity \cite{dN}. 
The explicit expressions for these 28-beins, in $SU(3)$-singlet sector,
in terms of four supergravity fields are
given in \cite{AW}. By restricting these to constants further, one gets
$SO(8)$ critical point and $SO(7)^{+}$ critical point. For the
$SO(7)^{+}$ flow, one has a single supergravity field which depends on
the radial coordinate of $AdS_4$ space.   

How does one determine the Killing vector?
On the seven-sphere, there exist eight scalar fields, $X^A(A=1, \cdots,
8)$ satisfying some constraints \cite{GRW,dWNW,DNP}.  
Using the $\Gamma$ matrices that are $SO(8)$ generators \cite{AW01,AI}, the Killing
vectors on the unit `round' ${\bf S}^7$,  that depend on the 7-dimensional
curved space $y^m$ via $X^A$, are given by \cite{dWNW}
\bea
\Kop_m^{\,\,IJ}(y) = \frac{1}{2} \, (\Gamma^{IJ})_{AB} \, \left( X^A \, \pa_m X^B -X^B \,
\pa_m  X^A \right). 
\label{K}
\eea
The two Killing vectors, 
$\Kop_m^{\,\,IJ}$ and $\Kop_m^{\,\,AB}$, are related to each other by
triality
where $\Kop_m^{\,\,IJ} =(\Gamma^{IJ})_{AB}\, \Kop_m^{\,\,AB}$.
The $28$ Killing vectors on seven-sphere can be expressed via the
Killing 
spinors satisfying the eight Killing spinor equations \cite{DNP}.
The 7-dimensional coordinates $y^m=(\th_1, \th_2, \th_3, \th_4,
\th_5, \th_6, \th)$ are related to 
the ${\bf R}^8$ coordinates $X^A(y^m)$ that are as follows \cite{AI}:
\bea
X^1(y^m) & = & \sin \th_1 \, \sin \th_2 \, \sin \th_3 \, \sin \th_4 \, \sin
\th_5 \, \sin \th_6 \, \sin \th, 
\nonu \\
X^2(y^m) & = & \cos \th_1 \, \sin \th_2 \, \sin \th_3 \, \sin \th_4 \, \sin
\th_5 \, \sin \th_6 \, \sin \th, 
\nonu \\
X^3(y^m) & = & \cos \th_2 \, \sin \th_3 \, \sin \th_4 \, \sin
\th_5 \, \sin \th_6 \, \sin \th, 
\nonu \\
X^4(y^m) & = & \cos \th_3 \, \sin \th_4 \, \sin
\th_5 \, \sin \th_6 \, \sin \th, 
\nonu \\
X^5(y^m) & = & \cos \th_4 \, \sin
\th_5 \, \sin \th_6 \, \sin \th, 
\nonu \\
X^6(y^m) & = & \cos \th_5 \, \sin \th_6 \, \sin \th, 
\nonu \\
X^7(y^m) & = & \cos \th, \nonu \\
X^8(y^m) & = & \cos \th_6 \, \sin \th.
\label{rect}
\eea
We denote the 11-th coordinate as the angle $\theta$.
Sometimes the Killing vectors can be written in terms of rectangular
coordinates $X^A$'s only by multiplying the transformation matrix between 
$X^A$ and $y^m$ into (\ref{K}), as in \cite{AI02}($\pa_m$ goes into $\pa_C$). 
Note $\sum_{A=1}^8 (X^A)^2=1$. For the round seven-sphere with radius $L$,
one should add $L$ in front of (\ref{K}). 

Then what is the generalized vielbein we mentioned before?
Let us take the contraction of $SO(8)$ indices present in (\ref{w})
and (\ref{K})
with upper index $m$  of Killing vector 
as follows \cite{dN87}:
\bea
e^m_{ij}(x, y) \equiv w_{ij}^{\,\,IJ}(x) \, \Kop^{mIJ}(y).
\label{emij}
\eea
Here the upper index $n$ of Killing vector
can be lowered via the `round' seven-sphere metric 
$\Gop_{mn}$ as
$\Kop_m^{\,\,IJ} = \Gop_{mn} \, \Kop^{nIJ}$.
Similarly, one has 
$\Kop^{mIJ} = \Gop^{mn} \, \Kop_n^{\,\,IJ}$ with its inverse metric 
$\Gop^{mn}$.
The $y$-dependence in (\ref{emij}) arises through 
the Killing vector while 4-dimensional data comes from 
$ w_{ij}^{\,\,IJ}(x)$ we defined in (\ref{w}).
This generalized vielbein satisfies the $SU(8)$ covariant  
``Clifford property'' \cite{dN85}
\bea
e^m_{ij} \, e^{njk} + e^{n}_{ij} \, e^{mjk} = \frac{1}{4} \, \delta_i^k \, e^{m}_{jl}
\, e^{nlj} = -8 \, \delta_i^k \,
\Delta^{-1} \, g^{mn}, 
\label{Cliff}
\eea
where the numerical factor $8$ depends on the normalization of Killing
vector we use in (\ref{K}).  In the convention of \cite{dN87}, this
coefficient becomes $2$ rather than $8$.
This determines the 7-dimensional metric $g_{mn}(x,y)$. 
Furthermore, the condition (\ref{Cliff}) is not satisfied all the time for
given Killing vector (\ref{K}) and one should find out the correct
Killing vectors which will satisfy (\ref{Cliff}),
using the $SO(8)$ invariance for the rectangular coordinates $X^A$. 
The warp factor is defined by
\bea
\Delta(x, y) \equiv
\sqrt{\frac{\mbox{det} \, g_{mn}(x, y) }{\mbox{det} \, \Gop_{mn}(y)}}.
\label{warp}
\eea
The last equation in (\ref{Cliff}) comes from the nonlinear metric ansatz
developed by \cite{dWNW}. Note the presence
of extra minus sign there due to the antisymmetric property of the
$SU(8)$  indices of generalized vielbein (\ref{emij}). 

The supersymmetry transformation of spin $\frac{1}{2}$ or
$\frac{3}{2}$ 
fermionic sector \cite{dN85,dN86}
contains
${\cal A}_m^{ABCD}(x,y)$ 
self-dual tensor with ``curved'' $SU(8)$ indices $A, \cdots, D$ and
curved 7-dimensional index $m$. 
Its ``flat'' version with flat $SU(8)$ indices $i, j, \cdots$ 
obtained by both multiplying 
the product of four Killing spinors $\eta^i_A \eta^j_B \eta^k_C \eta^l_D$ 
and contracting those curved indices, 
is given by 
\bea
{\cal A}_m^{ijkl} = \frac{4}{7}  m_7  \Kop_m^{KL} (
v^{ijLM}  u^{kl}_{KM} - u^{ij}_{LM}  v^{klKM})
-\frac{3}{28} \Dop_m  \Kop^{n[KL}  \Kop_n^{MN]}
( u^{ij}_{KL}  u^{kl}_{MN} - v^{ijKL}  v^{klMN}),
\label{calA}
\eea
where the covariant derivative $ \Dop_m$ contains the affine connection
as well as the ordinary partial derivative \cite{DNP}. 
The relative coefficients $\frac{4}{7}$ and $\frac{3}{28}$ were fixed
completely by 1) solving the generalized vielbein postulates(which
generalize the usual vielbein postulate of Riemannian geometry to the
complex geometry) 
and 2) requiring that  the T-tensor identified from 11-dimensional
supergravity also become $y$-independent \cite{dN87}. 
The $|m_7|$ is the inverse radius of seven-sphere.
This tensor will play the crucial role for the full nonlinear 4-forms
ansatz together with the generalized vielbein.
Note that the Killing vectors in the second term are contracted each
other with 7-dimensional index and one can lower the upper index $n$ 
by using the round metric as before.  
The $x$-dependence arises via $u,v$ 28-beins and $y$-dependence
appears in the Killing vectors and the covariant derivative acting on them.

On the other hand, the above ${\cal A}_m^{ijkl}$ tensor 
can be written in terms of 4-forms with internal 7-dimensional indices
explicitly \cite{dN85,dN86}.
By $SU(8)$ invariance, one can take a particular $SU(8)$ rotation as
in \cite{dN85}. Through the generalized vielbein postulate given 
in \cite{dN85,dN86},
we multiply a five-fold product of the generalized vielbein 
$e^m_{ij}$ (\ref{emij}) into (\ref{calA}) in order to preserve the $SU(8)$ covariance.
Using the various $\Gamma$ matrix properties in \cite{dN86}, one obtains
the nonlinear expression for the field strength given by \cite{dN87}
\bea
 \frac{4}{7} \, i \, f \, g_{n[p} \, \delta_{q]}^{m} +
  \frac{1}{2} F^{m}_{\,\,\,\,\, npq}  
= \frac{i}{480} \, \sqrt{2} \,
\Delta^4 \, \sqrt{\Gop} \, \varepsilon_{pqrstuv}  \, 
e^m_{ij} \, \left( e^{r}_{kk'} \, e^{sk'l'} \, e^t_{l'm'} \, e^{um'n'}
  \, e^{v}_{n'l} 
\right) \, 
{\cal A}_n^{ijkl},
\label{field}
\eea
where
the field strengths with
4-dimensional flat indices $\al, \cdots, \de$ appear in 
\bea
f \equiv \frac{1}{24 i} \,
\varepsilon^{\alpha \beta \gamma \delta} \, F_{\alpha \beta \gamma
  \delta}
= \frac{1}{24 i \Delta } \,
\varepsilon^{\mu \nu \rho \sigma} \, F_{\mu \nu \rho \sigma}.
\label{f}
\eea
Note that there exists a typo in \cite{dN87} and the factor $
\sqrt{\Gop}$ should be in the numerator not denominator(we fix here).
The $\Gop$ is the determinant of the round metric of seven-sphere.
In the right hand side of (\ref{field}), one sees the $SU(8)$
covariance in the product of between the five generalized vielbeins
and ${\cal A}_n^{ijkl}$ tensor. Moreover, the remaining $SU(8)$
indices are contracted with those of generalized vielbien and then
this leads to the $SU(8)$ invariance. 
It is amazing that the right hand side can decompose into the 
two tensors of the left hand side.
One can rewrite (\ref{field}) by lowering the upper index with 7-dimensional
metric  as follows:
\bea
 \frac{4}{7}  i  f   \left( g_{np}  g_{mq} -
g_{nq}  g_{mp} \right)+
   F_{mnpq}  
= 
\frac{i}{480} \, \sqrt{2} \, 
\Delta^4 \, \sqrt{\Gop} \, \eta_{pqrstuv} \, 
g_{m m'} \,
e^{m'}_{ij} \, \left( e^{r} \, e^s \, e^t \, e^u \, e^{v} \right)_{kl} \, 
{\cal A}_n^{ijkl},
\label{final}
\eea
where the 7-dimensional eta tensors with lower indices
$\eta_{pqrstuv}$ 
are purely numerical. We use a simplified notation for the
$kl$-element in the right hand side and the explicit components for
five-fold product are given in (\ref{field}) where there exists a
complex conjugation between the upper indices and lower ones for
$SU(8)$ we described in (\ref{w}). For the three cases we are
considering in this paper, the 28-beins are real and there is no
difference between the upper and lower indices appeaing in the
generalized vielbeins.   
Since the 7-dimensional metric we will use contains the factor
$\sqrt{\Delta}$, the warp factor-dependence of left hand side of
(\ref{final}) 
disappears when we use the second expression for $f$ in (\ref{f}).  

Therefore, the internal 4-form field strengths $F_{mnpq}$ with 7-dimensional indices  
are determined by the following six quantities:

1) the 
28-beins $u, v$ that appear in (\ref{w}), (\ref{emij}) and (\ref{calA}),  

2) the Killing vectors (\ref{K}) that are present in (\ref{emij}) and (\ref{calA})
on  the round seven-sphere, 

3) the covariant
derivative acting on these Killing vectors via (\ref{calA}), 

4) the warp
factor (\ref{warp}), 

5) the field strengths with
4-dimensional flat indices (\ref{f}) and 

6) the 7-dimensional metric. 

Now we compute all these quantities appearing in the right hand side
(\ref{final})
and compare the resulting expressions  with the left hand side of (\ref{final}).
Then one can read off the correct informations on the 4-forms which
will be the 11-dimensional solutions in the background we are
interested in. Since we already expressed the Killing vectors in terms
of 7-dimensional curved coordinates $y^m$ rather than the rectangular
coordinates $X^A$, the results will do not contain the rectangular
coordinates
and we do not have to do extra works, contrary to  the case of 
full nonlinear metric ansatz as in \cite{AI02}.  

\subsection{The mixed 4-form field strengths}

What about the other components for the 4-form field strengths?
For example, mixed 4-form field strengths with some internal indices
and some non-internal indices.
Although these are mentioned in \cite{dN87} at the end of paper, 
the explicit expressions are not known so far.
The full nonlinear expressions for the remaining 4-form field
strengths
(for the 4-forms $F_{\mu \nu m n}$ with two internal indices we will
describe at the end of this subsection)
can be obtained by projecting out 
the appropriate components in 
${\cal A}_{\mu}^{ijkl}$ using the four-dimensional 
results for these, as done exactly in \cite{dN87}.
The supersymmetry transformation of spin $\frac{1}{2}$ fermion sector
has
${\cal A}_{\mu}^{ijkl}$ tensor which is fully antisymmetric and
self-dual in the indices $i,j,k,l$ and
it is given by
\bea
{\cal A}_{\mu}^{ijkl} = -\frac{1}{48} \sqrt{2} \,e_{\mu}^{\alpha}
\, \varepsilon^{\alpha \beta \gamma \delta} \, F_{a \beta \gamma \delta}
\, \Gamma^b_{[ij} \, \Gamma_{kl]}^{ab} + \frac{1}{8} \sqrt{2} \, e_{\mu}^{\alpha}
\, F_{abc\alpha} \, \Gamma_{[ij}^{a} \, \Gamma_{kl]}^{bc} + \cdots,
\label{muijkl}
\eea
where the abbrebiated part of (\ref{muijkl}) contains
the term $ \Gamma_{[ij}^{a} \, \Gamma_{kl]}^{b}$. When we compute the six
generalized vielbein and  ${\cal A}_{\mu}^{ijkl}$ tensor as in
(\ref{field}), one should use the explicit form for the generalized
vielbein.
For example, the equation $(2.19)$ of \cite{dN87}. Then there exist
six Gamma matrices. Each Gamma matrix from each generalized vielbein.
The five of them can be reduced to two from the
identity given in \cite{dN86}.  Totally, one has three Gamma matrices 
$\Gamma_{ij}^{c} \, \Gamma_{kl}^{de}$ from six generalized vielbeins.
Then by combining these two factors, one can check that 
the quantity $ \Gamma_{[ij}^{a} \, \Gamma_{kl]}^{b} \,  
\Gamma_{ij}^{c} \, \Gamma_{kl}^{de}$ vanishes identically.
This feature is the same as the one in (\ref{field}).

One can project out the mixed 4-forms 
$ F_{a \beta \gamma \delta}$ and $ F_{abc\alpha}$ 
in $SU(8)$ invariant way as before. 
This leads to the following expression
\bea
e^m_{ij} \, \left( e^{[n} \, e^p \, e^q \, e^r \, e^{s]} \right)_{kl} \, 
{\cal A}_{\mu}^{ijkl} = i \, \sqrt{2} \, \Delta^{-4} \,
\frac{\varepsilon^{npqrstu}}{\sqrt{\Gop}}
\left( \frac{1}{3 \, \Delta}   \, \varepsilon_{\mu \nu\rho \sigma} \,
  F^{\nu\rho\sigma}_
{\,\,\,\,\,\,\,\,\,\, [t} \, \delta_{u]}^{m} +
  2 F^{m}_{\,\,\,\,\, \mu tu} \right),
\label{mid}
\eea
where we use some Gamma matrix identities again. 
Let us emphasize that in the original paper \cite{dN87}, 
they used the fact that the identity $\Gamma^{npqrs}_{CD} =
\frac{i}{2} \varepsilon^{npqrstu} \, (\Gamma_{tu})_{CD} \,
\frac{1}{\sqrt{g}}$ 
holds where $g$ is the determinant of the 7-dimensional metric. 
Via vielbeins 
$e_m^a$ and $e_{\mu}^{\alpha}$, 
the 4-forms  $F_{a \beta \gamma \delta}$ and $F_{abc\alpha}$
given in (\ref{muijkl}) are changed into $ F_{m \nu\rho\sigma}$ and 
$F_{\mu npq}$ (\ref{mid}) respectively.
In order to extract the 4-form field strengths, 
one can further simplify (\ref{mid}), by inverting it, as 
\bea
&& \frac{1}{3 \, \Delta}   \, \varepsilon_{\mu \nu\rho \sigma} \,
  \left( F^{\nu\rho\sigma}_
{\,\,\,\,\,\,\,\,\,\, p} \, g_{qm} - F^{\nu\rho\sigma}_
{\,\,\,\,\,\,\,\,\,\, q} \, g_{pm} \right) -
  2 F_{\mu m  pq} \nonu \\
&& = \frac{i}{480} \, \sqrt{2} \, 
\Delta^4 \, \sqrt{\Gop} \, \eta_{pqrstuv} \, 
g_{m m'} \,
e^{m'}_{ij} \, \left( e^{r} \, e^s \, e^t \, e^u \, e^{v} \right)_{kl}
\, 
{\cal A}_{\mu}^{ijkl}.
\label{final1}
\eea
Once we figure out the equation 
(\ref{final}), it is straightforward to compute this quantity also.
For example, 
the $kl$-component of five-fold product of generalized vielbein,  
where the explicit structure of indices are given in (\ref{field}), 
appears in (\ref{final1}) again and we do not have to compute this repeatedly.
Note that for both $G_2$ invariant flow and $SU(3) \times U(1)_R$
flow, 
it is known that 
the 4-forms appearing in the left hand side of (\ref{final1})
occur naturally. 
On the other hand, the 
${\cal A}_{\mu}^{ijkl}(x)$ tensor in (\ref{final1}) appears in the scalar
kinetic terms of 4-dimensional ${\cal N}=8$ gauged supergravity 
and it is given by
\bea
{\cal A}_{\mu}^{ijkl}(x) = -2\sqrt{2} \left( u^{ij}_{\,\,IJ} \, \pa_{\mu}
  v^{klIJ} 
- v^{ijIJ} \, \pa_{\mu} u^{kl}_{\,\,IJ}   \right).
\label{calAijkl}
\eea
In order to compute this tensor one has to know the $x$-dependence on
$u, v$ 28-beins. Since we are interested in the domain wall solutions,
one should have the first order differential equations between the
supergravity fields. These are found in \cite{AW} for $SU(3)$-singlet sector.  
For $SO(7)^{+}$ flow case we consider in this paper, 
the corresponding first order differential
equations are found in \cite{Ahn0812}.

Therefore, the mixed 4-form field strengths $ F_{a \beta \gamma
  \delta}$(or $F_{m \nu \rho \sigma}$ with world indices) and 
$ F_{\alpha bcd}$(or $F_{\mu npq}$ with world indices)     
are determined by the following four quantities:

1) the 
28-beins $u, v$ that appear in (\ref{w}), (\ref{emij}) and (\ref{calAijkl}),  

2) the Killing vectors (\ref{K}) that are present in (\ref{emij}) and (\ref{calA})
on  the round seven-sphere, 

3) the warp
factor (\ref{warp}) and 

4) the 7-dimensional metric. 

So far, we have considered the 4-forms, $F_{\mu\nu\rho\sigma}$ with no
internal indices and $F_{mnpq}$ with all the internal indices
in (\ref{final}),
$F_{m \nu\rho\sigma}$ with one internal index and $F_{\mu npq}$ with three
internal indices in (\ref{final1}). 
What happens for $F_{\mu \nu m n}$ with two internal indices?
According to the result of \cite{dN87}, there exists a relation 
$(u_{ij}^{\,\,IJ} + v_{ijIJ}) \, \overline{F}_{\mu
  \nu}^{-ij} = [F_{\mu \nu}^{\,\, IJ}]_{-}$ where the
$SO(8)$
field strength $F_{\mu \nu}^{\,\, IJ}$ is $F_{\mu \nu}^{\,\, IJ} =
\pa_{\mu} A_{\nu}^{\,IJ} - \pa_{\nu} A_{\mu}^{\, IJ} - 2 g A_{\mu}^{\,
K[I} \, A_{\nu}^{\, J] K}$ and $[F_{\mu \nu}^{\,\,
  IJ}]_{-}$ is the anti-self dual part of this field strength.
If there are no gauge fields in the 4-dimensional ${\cal N}=8$ 
gauged supergravity(the action consists of scalar and gravity part), 
then the above relation implies $\overline{F}_{\mu
  \nu}^{-ij}=0$(or $\overline{F}_{\alpha
  \beta}^{-ij}=0$). Furthermore, using the two spinors one can
contract the indices $i,j$ and this leads to ${\cal C}_{\alpha
  \beta}^{-AB}$ which is contained in $F_{\mu \nu m n}$. See the
equation (8.6) of \cite{dN86} where the multiple product of $e^m_{AB}$
tensor(sechsundfunfzigbein) are contracted with ${\cal C}_{\mu
  \nu}^{-AB}$. In other words, 
$F_{\mu \nu mn} \simeq \sqrt{g} \, \eta_{mnpqrst} \,
e^{pBC} \, e^q_{CD} \, e^{rDE} \, e^s_{EF} \, e^{tFA} \, \Delta^2
{\cal C}_
{\mu  \nu}^{-AB}$. Therefore,     
for the 11-dimensional background with domain wall we are considering, 
there exists no $F_{\mu \nu m n}$ with two internal indices.

However, in the context of AdS/CMT where the gauge fields of
4-dimensional ${\cal N}=8$ gauged supergravity play an
important role \cite{BHPW10}, it is necessary to obtain nonzero  $F_{\mu \nu m n}$
with two internal indices. See the relevant work by \cite{DH,KPT} where 
the supergravity theory is not realized by 4-dimensional gauged ${\cal
N}=8$ supergravity but the 4-forms with two internal indices and two
from 4-dimensional indices are nonvanishing due to the nonzero 2-form 
field strength along the 2-dimensions inside the 4-dimensions.  
It is an open problem to find out nontrivial $F_{\mu \nu m n}$ with 
two internal indices in this background in the context of
4-dimensional gauged ${\cal N}=8$ supergravity.

\section{The eleven-dimensional solutions}

In this section, 
at first, 
we compute the right hand side of (\ref{final}) for the known critical
points($SO(8)$ and $SO(7)^{+}$) 
by collecting (\ref{emij}), (\ref{calA}) with 11-dimensional metric 
explicitly
and compare them with the left hand side of (\ref{final}), for given
7-dimensional metric.
In other words, the quantities $f, g_{mn}$ and $F_{mnpq}$ are known for the
critical points. What we are doing newly 
is to calculate the right hand side of (\ref{final})
based on the six quantities we mentioned before
and to check whether the full nonlinear ansatz is right or not for
consistency check.
This is never done before and we will present the details in this section.

Later, we will consider membrane flow solution connecting between the $SO(8)$
critical point and the $SO(7)^{+}$ critical point. More precisely 
the flow solution contains $SO(8)$ critical point but does not contain 
$SO(7)^{+}$ critical point. One should use the
equation (\ref{final1}) also as well as (\ref{final}).    
The full nonlinear expressions  for the 4-form field strengths 
(\ref{final}) and mixed-form field strengths (\ref{final1})
will provide how to
obtain the new supersymmetric(or nonsupersymmetric) membrane flows 
in 11-dimensions, once the 4-dimensional RG flow equations  where the
supergravity fields vary with the radial coordinate of $AdS_4$ space 
are known.

\subsection{The ${\cal N}=8$ $SO(8)$ critical point}

Let us consider the $SO(8)$ critical point. The verification for this
critical point was done in \cite{dN87} already and it is a good exercise to
check this first.
For the round seven-sphere metric
\bea
g_{mn} = 
\left(\begin{array}{ccccccc}
s_{\th}^2 \, s_2^2 \, s_3^2 \, s_4^2 \, s_5^2 \, s_6^2 &0 &0 & 0& 0&0 &0  \\ 
0 & s_{\th}^2 \, s_3^2 \, s_4^2 \, s_5^2 \, s_6^2 & 0 & 0 & 0 & 0 &0 \\
0 & 0 & s_{\th}^2
 \, s_4^2 \, s_5^2 \, s_6^2 & 0 & 0 & 0 & 0 \\
0 & 0 & 0 & s_{\th}^2 \, s_5^2 \, s_6^2 & 0 & 0 & 0 \\
0 & 0 & 0 & 0 & s_{\th}^2 \, s_6^2 & 0 & 0\\
0 & 0 & 0 & 0 & 0 & s_{\th}^2 & 0 \\
0 & 0 & 0 & 0 & 0 & 0 &  1
\end{array} 
\right) = \Gop_{mn},
\label{roundmetric}
\eea
one writes down the square root of determinant as 
\bea
\sqrt{\Gop} = s_{\th}^6 \, s_{2} \, s_{3}^2 \, s_{4}^3 \,
s_{5}^4 \, s_{6}^5, \qquad s_{\th} \equiv \sin \th, \qquad s_i \equiv
\sin \th_i. 
\label{det}
\eea
From the definition of (\ref{warp}), the warp factor $\Delta$ becomes
$1$ since $g_{mn} =\Gop_{mn}$.
The generalized vielbein $e^m_{ij}$ is given by (\ref{emij}) with $w_{ij}^{\,\,IJ} = 
\delta_{ij}^{\,\,IJ}$ and the Killing vector can be obtained from (\ref{K})
and (\ref{rect}) explicitly.
The ${\cal A}_m^{ijkl}$ tensor 
is given by (\ref{calA}) where the first term
vanishes because $v^{ijKL} =0$ for $SO(8)$ critical point. 
In order to compute the second term of 
(\ref{calA}), it is better to use some property of Killing vector when
acting on the covariant derivative. That is, one can rewrite as 
$ \Dop_m  \Kop^{nKL}  \Kop_n^{MN} =  
(\Dop_m  \Gop^{nn'}) 
\Kop_{n'}^{KL}  \Kop_n^{MN} + \Gop^{nn'} (\Dop_m  \Kop_{n'}^{KL})
\Kop_n^{MN}+
\Gop^{nn'} \Kop_{n'}^{KL}  (\Dop_m \Kop_n^{MN})$ with the metric of 
round seven-sphere in terms of three terms. The first term vanishes and
the remaining terms can be simplified further. Then this 
becomes 
\bea
 \Dop_m  \Kop^{nKL}  \Kop_n^{MN}=
m_7  \left( \Gop^{nn'}   \Kop_{m}^{K'[K} \Kop_{n'}^{L]K'}
\Kop_n^{MN} +   \Gop^{nn'} \Kop_{n'}^{KL} \Kop_m^{M'[M}
\Kop_{n}^{N]M'}\right),
\label{dksq}
\eea
where we used the relation $(A.3)$ of \cite{dN87} together with 
Killing spinor equations and Killing vectors expressed in terms of 
the Killing spinors of \cite{dN87} and $\Dop_m  \Gop^{nn'}=0$.
Of course, there appears the minus sign  in (\ref{dksq}) for the skew-whiffing or
orientation reversal of seven-manifold \cite{DNP}.

By substituting all of these (\ref{roundmetric}), (\ref{det}) 
including (\ref{dksq}) 
into the right hand side of (\ref{final}), one
arrives at the final nonzero components and they are given in the
Appendix $A$ explicitly. By
computing the left hand side of (\ref{final}) with the condition
$g_{mn} =\Gop_{mn}$ (\ref{roundmetric}), 
one concludes that the following relations for the 4-form field
(\ref{f}) and the internal ones should hold
\bea
f = 3 \sqrt{2} \, m_7, 
\qquad F_{mnpq} =0,
\label{solutionround}
\eea
which was observed in \cite{dN87} also. This is well-known
Freund-Rubin solution
for round seven-sphere compatification \cite{FR}.
In the appropriate normalization, (\ref{solutionround}) implies that 
the 4-form along the membrane  is given by $F_{1234}= 3 m_7 \,
e^{3A(r)}$ in the background 
\bea
ds_{11}^2 =\Delta^{-1} \, \left(dr^2 +e^{2 A(r)}
\, \eta_{\mu\nu}\, dx^\mu dx^\nu \right)+ ds_7^2,
\label{11d}
\eea
where the 3-dimensional metric is $\eta_{\mu \nu}=(-, +, +)$, the
radial coordinate is transverse to the domain wall, and the scale factor
$A(r)$ in (\ref{11d}) behaves linearly in $r(\equiv x^4)$ 
at UV and IR regions. The warp factor is defined as (\ref{warp}). 
Of course, at the $SO(8)$ critical point, the 28 beins $u$ is constant and
$v$ is equal to zero and (\ref{calAijkl}) also vanishes. From
(\ref{final1}),
there are no mixed 4-forms.

\subsection{The nonsupersymmetric $SO(7)^{+}$  critical point}

Let us describe the next nontrivial example.
If one uses the previous rectangular coordinates 
(\ref{rect}), then the Clifford property 
(\ref{Cliff}) does not satisfy.
For this critical point, one should transform the rectangular
coordinates (\ref{rect}) using transformation matrix $R$ \cite{AI02}
as follows:
\bea
\widetilde{X} = R^{-1} \, X,
\label{newX}
\eea
where the $8 \times 8$ orthogonal matrix $R$ in (\ref{newX}) is given by
\bea
R= \left(\begin{array}{cccccccc} 
0 & 0 & 0& 0&0 &0 &-1 & 0 \\ 
0 & 1 & 0& 0&0 &0 &0 &0  \\
-\frac{1}{\sqrt{2}} & 0 &0 &0 & \frac{1}{\sqrt{2}}&0 &0 &0  \\ 
0 & 0 &0 &-\frac{1}{\sqrt{2}} & 0&0 &0 & \frac{1}{\sqrt{2}}  \\ 
0 & 0 &-1 &0 & 0&0 &0 &0  \\ 
0 & 0 &0 &0 & 0&1 &0 &0  \\
-\frac{1}{\sqrt{2}} & 0 &0 &0 & -\frac{1}{\sqrt{2}}&0 &0 &0  \\  
0 & 0 &0 &-\frac{1}{\sqrt{2}} & 0 &0 &0 & -\frac{1}{\sqrt{2}}  \\ 
\end{array} \right). 
\label{Rmatrix}
\eea
The reason for this is due to the fact that 
the generalized vielbein for $SO(7)^{+}$ critical point 
should also satisfy the Clifford property (\ref{Cliff}).
This is a useful check whether one has the right choice for the
Killing vectors.
Originally, the presence of $R$ in (\ref{Rmatrix}) 
was necessary in order to obtain the standard Kahler form from the
inverse metric for $SU(4)^{-}$ critical point in the
context of full nonlinear metric ansatz \cite{dWNW}.

The 7-dimensional metric is given by
\bea
g_{mn} = \sqrt{\Delta \, a} \,
\left(\begin{array}{ccccccc}
s_{\th}^2 \, s_2^2 \, s_3^2 \, s_4^2 \, s_5^2 \, s_6^2 &0 &0 & 0& 0&0 &0  \\ 
0 & s_{\th}^2 \, s_3^2 \, s_4^2 \, s_5^2 \, s_6^2 & 0 & 0 & 0 & 0 &0 \\
0 & 0 & s_{\th}^2
 \, s_4^2 \, s_5^2 \, s_6^2 & 0 & 0 & 0 & 0 \\
0 & 0 & 0 & s_{\th}^2 \, s_5^2 \, s_6^2 & 0 & 0 & 0 \\
0 & 0 & 0 & 0 & s_{\th}^2 \, s_6^2 & 0 & 0\\
0 & 0 & 0 & 0 & 0 & s_{\th}^2 & 0 \\
0 & 0 & 0 & 0 & 0 & 0 &  \frac{\xi^2}{a^2}
\end{array} 
\right),
\label{7dmetric}
\eea
where the deformed norm in (\ref{7dmetric}) is given by
\bea
\xi^2 = a^2 \, \cos^2 \th + b^2 \, \sin^2 \th.
\label{xi}
\eea
One can also express this 7-dimensional metric in terms of rectangular
coordinates (\ref{rect}) and the eccentricity of 7-dimensional 
ellipsoid  depends on $(a,b)$.
The warp factor (\ref{warp}), together with (\ref{xi}),  becomes
\bea
\Delta = a^{-1} \, \xi^{-\frac{4}{3}}.
\label{warp1}
\eea
Of course, for round seven-sphere, we have $a=b=1$(and $\xi^2=1$) and
the metric (\ref{7dmetric}) becomes the round metric (\ref{roundmetric}).
The geometric parameters $(a,b)$ 
in the 7-dimensional ellipsoid can be identified with the $AdS_4$
supergravity fields.  
How does one see the $SO(7)$ symmetry? By writing the 7-dimensional 
warped ellipsoid as 
$ds_7^2 = \sqrt{\Delta \, a } \left( \frac{\xi^2}{a^2} d \theta^2 + \sin^2
  \theta \, d\Omega_6^2 \right)$, one sees that the metric contains 
six-sphere whose isometry is nothing but $SO(7)$.
Here the $SO(7)^{+}$-invariant critical point fixes
the $AdS_4$ supergravity fields as follows:
\bea
a = 5^{\frac{1}{4}}, \qquad b = 5^{-\frac{1}{4}}.
\label{vev}
\eea
The scalar potential is a function of two supergravity fields and the
$SO(7)^{+}$
symmetry further restricts to them. The derivative of scalar potential
with respect to a single supergravtiy field vanishes at the critical point.
However, the derivative of superpotential at the critical point does
not vanish.
The warp factor (\ref{warp1}) becomes
$ \Delta = 
\frac{5^{\frac{1}{12}}}{(3+ 2 \cos 2\theta)^{\frac{2}{3}}}$ by
substituting (\ref{vev}) into (\ref{warp1}) and (\ref{xi}).
In next subsection, we will consider the case where $AdS_4$ supergravity fields 
$(a,b)$ vary with the radius $r$ of $AdS_4$ space in the
11-dimensional background (\ref{11d}).
The $SO(7)^{-}$ critical point corresponds to $a=b=\frac{\sqrt{5}}{2}$.
Due to the $\frac{\xi^2}{a^2}=1$ (\ref{xi}), there is no deformation in round
seven-sphere except the overall factor (\ref{7dmetric}).
Since the 28-bein $v$ is imaginary along the $SO(7)^{-}$ flow \cite{Ahn0812}, the
generalized vielbein has imaginary part as well as real part and this
makes the computations complicated.

Now we are ready to compute the right hand side of the 
equation (\ref{final}).
It is known in \cite{AW} that the 28-beins are given in terms of 
$AdS_4$ supergravity fields (\ref{vev}). The Killing vector is given
by (\ref{K}) where $X$ is replaced by $\widetilde{X}$ in (\ref{newX}).
The warp factor is given by (\ref{warp1}) with (\ref{xi}) and (\ref{vev}).
Finally, the 7-dimensional metric is given by (\ref{7dmetric}).
One can plug these data into the right hand side of (\ref{final}) and
it turns out that there exists a mismatch. It does not provide the
known 4-forms: nonzero constant $f$ with vanishing $F_{mnpq}$.

How does one resolve this problem? We have to look at what we have
done so far again. In order to do that,
let us introduce two real constants in front of each term of 
${\cal A}_m^{ijkl}$ in (\ref{calA}) $k_1$ and $k_2$ as follows:
\bea
\widetilde{\cal A}_m^{ijkl}(x,y) & = & 
 k_1 \, \frac{4}{7}  m_7  \Kop_m^{KL} (
v^{ijLM}  u^{kl}_{KM} - u^{ij}_{LM}  v^{klKM})
\nonu \\
& - & k_2 \, \frac{3}{28} \Dop_m  \Kop^{n[KL}  \Kop_n^{MN]}
( u^{ij}_{KL}  u^{kl}_{MN} - v^{ijKL}  v^{klMN}).
\label{newA}
\eea
We want to see whether we can fix these constants by requiring that 
the equation (\ref{final}) should satisfy for $SO(7)^{+}$ critical point.
Therefore, we compute the right hand side of (\ref{final})
where $\widetilde{\cal A}_m^{ijkl}$ (\ref{newA}) is used. 
It turns out that the two constants can be fixed as follows:
\bea
k_1 =\frac{1}{2}, \qquad k_2 =-\frac{1}{3}.
\label{k1k2}
\eea
Unfortunately, there are extra minus sign in the second term of
(\ref{calA})
and the numerical factors are not equal to each other. 
Are the numerical factors $\frac{4}{7}$ and $\frac{3}{28}$ in (\ref{calA}) wrong?

How does one understand this behavior?
Let us first consider the sign problem.
The minus sign in $k_2$ can be understood from the Killing spinor
equations. The way of appearance of $m_7$ in (\ref{dksq}) comes from
the Killing spinor equations. The third choice for Killing vector is
to take the same Killing vector (\ref{K}) with (\ref{newX}) 
but with spinors satisfying 
the Killing spinor equations of the opposite sign \cite{dWNW}.
This is equivalent to send the siebenbein $\eop_m^a$ to $-\eop_m^a$.    
Then the right hand side of (\ref{dksq}) has an extra minus sign.
Let us consider the different relative factors.
Now then how does one understand the relative coefficient $\frac{2}{3}$ between
$k_1$ and $k_2$ in (\ref{k1k2})? 
One way to see this is to introduce two real
parameters as follows:
\bea
g_{mn} \rightarrow l_1^2 \, g_{mn}, \qquad 
\Kop_m^{IJ} \rightarrow l_2^2 \, \Kop_m^{IJ}.
\label{trans}
\eea
According to (\ref{trans}), one knows how the inverse metric $g^{mn}$,
the Killing vector with upper index $\Kop^{mIJ}$, the generalized
vielbein $e^m_{ij}$, and $e_{mij}$ transform.
Then one can easily check that under the condition 
$\frac{l_2^4}{l_1^2}=\frac{2}{3}$, the tensor (\ref{newA}) will
provide the correct result.
In other words, by using the equation (\ref{final}) with modified
tensor (\ref{newA}) with (\ref{k1k2}), the right hand side is
summarized in the Appendix B. 
Then by reading off the left hand side, one gets 
it turns out that
\bea
f = \sqrt{2} \, m_7, 
\qquad F_{mnpq}=0.
\label{sol}
\eea
This is a solution
for ellipsoidal deformation of the 7-manifold \cite{dNso7}.
The solution (\ref{sol}) corresponds to the nonzero component of 4-form 
as $F_{1234}= 5^{\frac{3}{4}} m_7 \,
e^{3A(r)}$ in the background (\ref{11d}).
What happens for $a=b=1$ limit? It is easy to see that it 
reproduces the result of
subsection 3.1 as we expect.
We will consider what happens if we
turn on certain supergravity
field in the $AdS_4$ supergravity(i.e., $a(r)$) where it approaches to zero in the
UV and develops a nontrivial profile as a function of $r$ as one goes to 
the IR next subsection \footnote{Recently, in \cite{NP2011}, the
most  general solution of the generalized vielbien
postulate $(3.2)$ of \cite{dN87} (corresponding to $(2.14)$ of
\cite{NP2011}), by adding a homogeneous term which does not affect the
$T$ tensor of 4-dimensional gauged supergravity theory,  is
found. They also presented the complete expression for the flux. Let
us describe how our results can fit their flux lift
formulae. According to their $(6.73)$ of \cite{NP2011}, for
$\alpha_m^{\,\,n}=-\frac{5}{7} \delta_m^{\,\,n}$, one sees that the
coefficient from the first two terms in the right hand side 
becomes $-\frac{2}{7}$ and the coefficient from the remaining terms 
becomes $\frac{2}{28}$. By adding these correction terms to
(\ref{calA}), 
one has  $\frac{4}{7} +(-\frac{2}{7})=\frac{2}{7}$ and  $-\frac{3}{28}
+(\frac{2}{28})=-\frac{1}{28}$ respectively. This is consistent with
the results (\ref{k1k2}) with (\ref{newA}) because the former is equal to $k_1
\frac{4}{7}(=\frac{2}{7})$ and the latter is equal to $-k_2 \frac{3}{28}(=
\frac{1}{28})$ except the minus sign. As we described before, this
sign problem can
be resolved by using the Killing spinor equations of the opposite
sign. 
We also have checked the equation $(2.28)$ of \cite{NP2011} for
$SO(7)^{+}$ case and the internal 4-form flux $F_{mnpq}$ vanishes as
expected in (\ref{sol}).  }. 

For convenience, we present the explict expressions for the
generalized vielbeins in the Appendix C where the $a(r)$-dependence
appears and this holds for the three cases in this paper. For
$a(r)=1$, the expressions in the Appendix C will give those in the
subsection 3.1 and for $a(r)=5^{\frac{1}{4}}$, those correpsond to the
ones in the subsection 3.2 and finally for the general $a(r)$ with
domain wall condition, the equation in the 
Appendix C will provide the generalized
vielbein in next subsection.  

\subsection{The nonsupersymmetric $SO(7)^{+}$ invariant flow}

So far, we have described two cases, $SO(8)$ critical point and 
$SO(7)^{+}$ critical point. Now let us consider more general case.
The supergravity fields vary with 4-dimensional space-time $x$.
In particular, we are interested in the domain wall solutions.  
The first order differential equations from $SO(8)$ to $SO(7)^{+}$
are written as \cite{Ahn0812}
\bea
\frac{d a(r)}{d r} =-\frac{1}{2L} \sqrt{a(r)} \, [a(r)^4-1],
\qquad
\frac{d A(r)}{ d r} = \frac{1}{4L} \frac{a(r)^4+7}{\sqrt{a(r)}}.
\label{domain}
\eea
At the supersymmetric $SO(8)$ critical point(i.e., $a(r)=1$), the
right hand side of first equation (\ref{domain}) vanishes while
at the nonsupersymmetric $SO(7)^{+}$ critical point(i.e., $a(r) =
5^{\frac{1}{4}}$) those quantity does not vanish.
The $SO(8)$ gauge coupling constant $g$ of 4-dimensional gauged 
${\cal N}=8$ supergravity
is replaced with $\frac{\sqrt{2}}{L} (=\sqrt{2} m_7)$.

The 11-dimensional bosonic field equations are  \cite{CJS}
\bea
R_{M}^{\;\;\;N} & = & \frac{1}{3} \,F_{MPQR} F^{NPQR}
-\frac{1}{36} \de^{N}_{M} \,F_{PQRS} F^{PQRS},
\nonu \\
\nabla_M F^{MNPQ} & = & -\frac{1}{576} \,E \,\ep^{NPQRSTUVWXY}
F_{RSTU} F_{VWXY},
\label{fieldequations}
\eea
for given 11-dimensional metric (\ref{11d}) and (\ref{7dmetric}) with
$L^2$ factor  and 4-form field strengths.
The covariant derivative $\nabla_M$ 
on $F^{MNPQ}$ in 
(\ref{fieldequations})
is given by 
$E^{-1} \pa_M ( E F^{MNPQ} )$ together with elfbein determinant 
$E \equiv \sqrt{-g_{11}}$. The epsilon tensor 
 $\ep_{NPQRSTUVWXY}$ with lower indices is purely numerical.
Imposing the $r$-dependence to the vacuum expectation value $a(r)$, the
11-dimensional metric (\ref{11d}) generates the Ricci tensor
components \cite{AI}. Applying the RG flow equations (\ref{domain}), 
all the $r$-derivatives in the Ricci tensor components can be replaced
with polynomial of $a(r)$. 

Now we want to obtain 11-dimensional solution satisfying
(\ref{fieldequations}) under the RG flow equations. 
Let us first consider the mixed 4-form field strengths using the
equation (\ref{final1}).
For the $G_2$ invariant sector, we have relations 
$c(r)=a(r)$ and $d(r) =b(r)$ \cite{AI02}. 
Further constraint $b(r)=\frac{1}{a(r)}$
gives the $SO(7)^{+}$ invariant flow where the original field given in
\cite{Warner83} is related to $a(r) =
e^{\frac{\lambda(r)}{\sqrt{2}}}$.
The parametrization for the $SU(3)$-singlet space  \cite{Warner83,Warner83-1}
contains the complex self-dual tensor describing $35$ scalars and 
$35$ pseudo scalars of 4-dimensional
gauged ${\cal N}=8$ supergravity.  
The supergravity fields reduce to two by $G_2$ invariance and 
these two further reduce to one by $SO(7)^{+}$ symmetry.

Let us consider the equation (\ref{final1}). Due to the domain wall
solution (\ref{domain}), the nontrivial solution of (\ref{final1})
appears only when the $\mu$ index is equal to $r=x^4$.
For other case($\mu=1, 2, 3$), the right hand side of (\ref{final1})
vanishes and this implies that the 4-forms in the left hand side
should be equal to zero.  For fixed $\mu=4$, there exist three free
indices $m, p, q$.
Then one can compute the right hand side explicitly. Then it turns out 
the nonzero contributions arise when $(m,p,q)=(1,1,7)$,
$(2,2,7)$, $(3,3,7)$, $(4,4,7)$, 
$(5,5,7)$, $(6,6,7)$. These are given in the
(\ref{nonzero}) of the Appendix D. 
This implies that the antisymmetric 4-form $F_{4npq}$ vanishes because 
two of indices are equal from the right hand side of (\ref{final1}).
So the remaining 4-forms in the left hand side can be read off
directly.
By rescaling the ${\cal A}_{\mu}^{ijkl}$ tensor by $ \frac{i
\sqrt{2}}{ g}$, one obtains the nonzero mixed 4-form with the indices 
$(\mu,\nu,\rho,m)=(1,2,3,11)$
\bea
F_{123\, 11}(r,\theta) =-\frac{e^{3A(r)}}{2\sqrt{a(r)}} \left[ a(r)^4 -
1\right] \, \sin 2\theta.
\label{12311}
\eea
At $SO(8)$ critical point($a(r)=1$), this vanishes. At the $SO(7)^{+}$
critical point($a(r) = 5^{\frac{1}{4}}$), the above $F_{123\,11}$ (\ref{12311}) does
not vanish. This implies that the nonsupersymmetric $SO(7)^{+}$ invariant
flow solution does not include the previous $SO(7)^{+}$ critical point 
solution where there are no mixed 4-forms according to (\ref{sol}). 
This feature looks different from the ones for 
supersymmetric flow cases where 11-dimensional flow solutions contain
either $SU(3) \times U(1)_R$ critical point or $G_2$ critical point at
the IR fixed point. The reason comes from the domain wall solutions in 
(\ref{domain}). Since the right hand side of the first equation does
not vanish at the $SO(7)^{+}$ critical point, this nonzero effect will
go into the expression (\ref{calAijkl}) and eventually the equation 
(\ref{final1}). However, the supersymmetric cases give the vanishings
of derivatives of 
supergravity fields with respect to the $r$ at the critical point.
The overall factor $[a(r)^4 -1]$ in (\ref{12311}) comes from the
tensor (\ref{calAijkl}) which has the supergravity derivative with
respect to $r$. This derivative is replaced by the first equation of
(\ref{domain})
and the right hand side of it has this overall factor. 

It is not obvious to see the $\theta$-dependence of (\ref{12311}) from
the (\ref{final1}) because we do not see any this dependence from the
7-dimensional metric, the generalized vielbein, the tensor
(\ref{calAijkl}) or the five-fold product of generalized vielbein.
Only after all the summations over the contracted indices 
are completed, the $\theta$-dependence occurs.  

Let us move the 4-forms with internal space indices and the 4-forms
with the membrane indices.
As we have done for the $SO(7)^{+}$ critical point case, 
by using the equation (\ref{final}) with modified
tensor (\ref{newA}) with (\ref{k1k2}) along the RG flow, 
the right hand side is
summarized in the Appendix D. 
Therefore, the 4-forms do not change, compared with the $SO(7)^{+}$
critical point case and they are given in (\ref{sol}).
It seems that the vanishing of $F_{mnpq}$ is reasonable but the
constant $f = \sqrt{2} m_7$ is not what we want to have because it
does not tell us any $r$-dependence on the 4-form. See the results in
(\ref{so7flow}) of the Appendix D. In order to
generalize the ansatz to the flow solution also as well as the
critical points solutions we have described so far, one has to introduce
some $(r,\theta)$ dependent factor in the first terms of (\ref{final}). 
This extra piece can be determined by Einstein equation. Or the other
possibility comes from the presence of  the inverse of this extra piece as an
overall factor in the right hand side of (\ref{final}).  
An immediate question arises. If we make an replacement by $m_7
\rightarrow \widetilde{m}_7(r,\theta)$ on the full linear ansatz of \cite{dN87}, 
can we see the inverse of extra piece from ${\cal A}_{n}^{ijkl}$
tensor 
automatically? 
Maybe the known supersymmetric critical point and flow solutions will
help for us to analyze for the above ambiguity on whether the extra
structure should appear in the 4-form in the left hand side or the
right hand side of (\ref{final}) because we will have further
information on the nonzero internal 4-forms and this will provide some
implication behind the ansatz (\ref{final}).    

One can substitute the 4-form (\ref{12311}) into the Einstein equation 
(\ref{fieldequations}). The Ricci component $R_4^{\,11}$ provides 
the product of $F_{1234}$ and $F_{123\,11}$. Then 
one can obtain the 4-form $F_{1234}$ explicitly and it is 
\bea
F_{1234}(r,\theta) & = & \frac{e^{3A(r)}}{2 L \, a(r)} 
\left[ - a(r)^8 \, \cos^2 \theta +
a(r)^4 \, (4 + 3 \cos 2\theta) + 5 \sin^2 \theta \right]
\nonu \\
& = &  \frac{e^{3A(r)}}{2 L \, a(r)} \left[ a(r)^{\frac{1}{2}} \,
\Delta(r,\theta)^{-\frac{3}{2}} \left( 5- a(r)^{4} \right) + 2 a(r)^4 \right].
\label{1234}
\eea
Due to the factor $(5-a(r)^4)$, at the $SO(7)^{+}$ critical point,
this 4-form (\ref{1234}) reduces to the one considered in (\ref{sol})
because the first two terms vanish and the remaining term becomes 
$F_{1234} = \frac{5^{\frac{3}{4}}}{L} e^{3A(r)}$ in subsection 3.2. 
Of course, at $SO(8)$ critical point, this reduces to the previous 
$F_{1234} = \frac{3}{L} e^{3A(r)}$ in subsection 3.1. 
For convenience, we present the warp factor (\ref{warp1}) here
\bea
\Delta(r,\theta) = \frac{2^{\frac{2}{3}} \, a(r)^{\frac{1}{3}}}{[1+ a(r)^4
  +(-1 +a(r)^4) \, \cos 2\theta ]^{\frac{2}{3}}}. 
\label{warpdel}
\eea 

The 11-dimensional Bianchi identity $\pa_{[M} F_{NPQR]}=0$ can be
checked explicitly. From the particular component of $\pa_{[11} F_{1234]}=0$,
one obtains the relation $\pa_{\theta} F_{1234}-\pa_r F_{123 \,11}=0$
where the derivatives with respect to $x^{\mu}$($\mu =1, 2, 3$) are vanishing.
Then it is easy to see $\pa_{\theta} F_{1234}=\pa_r F_{123 \,11}$
by substituting the two results (\ref{12311}) and (\ref{1234}).
One can easily check that the other components of Bianchi identity
give the  trivial result.
We have an extension of the Freund-Rubin compactification and 
the 3-form gauge field with 3-dimensional membrane indices 
looks like $A_{123}(r,\theta) \sim e^{3A(r)} \,
\widetilde{W}(r,\theta)$
where $\widetilde{W}(r,\theta)$ is so-called geometric superpotential.
This quantity is 11-dimensional lift of 4-dimensional superpotential
in the sense that the former becomes the latter for particular fixed
internal angle.  
Note the $\theta$-dependence on the geometric superpotential here. 
One can check that there are no other components of 3-form gauge
fields having other directions due to the presence of nonzero 4-forms 
(\ref{1234}) and (\ref{12311}).

For the the Maxwell equations in (\ref{fieldequations}), one can also
check the two 4-forms (\ref{12311}) and (\ref{1234}) satisfy the
second equation of (\ref{fieldequations}). In particular, the $(npq),
(np \, 11)$
and $(np4)$-components of Maxwell equations are trivially satisfied.
In other words, the left hand side and the right hand side are identically 
zero. For the component $(\nu \rho \sigma)$ of Maxwell equations, 
the left hand side can be
written as
\bea
\pa_r \, \left[ \Delta(r,\theta)^3 \, \sin^6 \theta \, e^{-3A(r)}\, 
F_{1234}(r,\theta) \right]
+ \pa_\theta \, \left[ g_{44} \, g^{11\,11} \,  \Delta(r,\theta)^3 \, \sin^6
  \theta \,
  e^{-3A(r)}\, F_{123\,11}(r,\theta) \right].
\label{max}
\eea
Now one substitutes the explicit expressions for the nonzero 4-forms 
(\ref{1234}) and (\ref{12311})   
into this expression (\ref{max}) and it leads to vanish via (\ref{warpdel}) and
(\ref{domain}), together with (\ref{11d}) and (\ref{7dmetric}). 
On the other hands, the
right hand side of $(\nu \rho \sigma)$ component of Maxwell equations
also gives zero because there are no internal 4-form field strengths.

We also checked that 
the other way to compute (\ref{final}) is given the equation $(7.8)$
of \cite{dN87} and
those computations also give the consistent results.
For the computation $e^m_{ij} \, {\cal A}_n^{ijkl}$ in
(\ref{final}), 
one can use the $SU(8)$ covaraint derivative with $SU(8)$ connection
which is known.
Using 
${\cal D}_n \, 
e^m_{ij}=\Dop_n \, e^m_{ij} +{\cal B}_{n[i}^k \, e_{j] k}^m$, one can
read off the above $SU(8)$ covariant quantity in different way.
The $SU(8)$ connection is related to 
${\cal B}_{mij}^{\,\,\,\,\,\,\,\,kl}$ tensor through 
${\cal B}_{mij}^{\,\,\,\,\,\,\,\,kl}=\delta_{[i}^{\,\,[k} 
{\cal B}_{mj]}^{\,\,\,\,\,\,l]}$ which is given in \cite{dN87}
explicitly
as follows:
\bea
{\cal B}_{mij}^{\,\,\,\,\,\,\,\,kl} =
 \frac{4}{7}  m_7  \Kop_m^{IJ} (
u_{ij}^{JK}  u^{kl}_{IK} - v_{ijJK}  v^{klIK})
+\frac{3}{28} \Dop_m  \Kop^{n[KL}  \Kop_n^{MN]}
( u_{ij}^{[IJ}  v^{klKL]} - v_{ij[IJ}  u^{kl}_{KL]}).
\label{calB}
\eea
From (\ref{calB}), one gets the $SU(8)$ connection and computes the 
covariant derivative acting on the generalized vielbeins. 
At the $SO(8)$ critical point, the first term of (\ref{calB}) 
survives since $v^{ijKL}$ vanishes.

Other way to check that the 11-dimensional solution by (\ref{12311}) and
(\ref{1234}) 
is correct is to consider the previous 11-dimensional solution for
supersymmetric $G_2$ invariant flow. Since the group $SO(7)$ has its
subgroup $G_2$, at least the 11-dimensional solution for $SO(7)^{+}$
invariant flow preserves the $G_2$ symmetry in the 11-dimensional
metric and the 4-forms. As observed previously, the metric 
has an isometry of $SO(7)$. The coefficient functions appearing in the
4-forms of $G_2$ invariant flow occur in the 4-forms with
4-dimensional indices, 4-forms with three internal indices, and 4-forms
with four internal indices.  By examining these more closely, one
realizes that many of these coefficient functions contain $[a(r)b(r)-1]$
factor.
Therefore, as soon as we impose the $SO(7)^{+}$ constraint
$b(r)=\frac{1}{a(r)}$ into these coefficient functions, they vanish.
What remains for the 4-forms is exactly the components of $F^{AI}_{1234}$ and 
$F^{AI}_{1235}$ in
\cite{AI,AW10-1} where the 5-th direction is the direction of $\theta$.
We observe that they are the same as the ones (\ref{1234}) and
(\ref{12311}) exactly.

For the maximally supersymmetric 
$SO(8)$ limit on the $SO(7)^{+}$ invariant flow, the result
reproduces the one in subsection of 3.1 while the nonsupersymmetric 
$SO(7)^{+}$ limit on the same flow does not
give the result of subsection 3.2. 
This is one of the reasons why we analyze the subsection 3.2 separately.   
As long as the supergravity fields do not vary with respect to the
radial coordinate, then we can go to the subsections 3.1 or 3.2. As
they vary, this subsection holds along the RG flow which contains the
$SO(8)$ critical point.

We explicitly computed the 11-dimensional supersymmetry for the RG
flow in the Appendix E. There exists no supersymmetry except the
$SO(8)$ critical point which has a maximal supersymmetry by solving 
(\ref{newint}).

\section{Conclusions and outlook }

Using the de Wit and Nicolai's formula (\ref{final}), we computed the
right hand side explicitly for three cases 1) $SO(8)$ critical point,
2) $SO(7)^{+}$ critical point and 3) $SO(7)^{+}$ flow. 
For our simple Killing vectors and 7-dimensional metric, one should
use the modified tensor given by (\ref{newA}). For the last case, 
we should also consider a new formula (\ref{final1}) which appears in
this paper for the first time. 

So far we have considerd only some part of membrane flows.
The known supersymmetric membrane flows are given by $G_2$
invariant flow and $SU(3) \times U(1)_R$ invariant flow
\cite{CPW,AW10}. 
These are 11-dimensional lifts of 4-dimensional domain wall solutions
in \cite{AW10,AP}. One should
also observe these flows by using the methods in this paper 
based on (\ref{final}) and (\ref{final1}). 
We expect to have the nonzero 4-form field strengths $F_{mnpq}$ with
internal indices. This will provide how the two tensors appearing in
the left hand side of (\ref{final}) decompose nontrivially. 
The main difficulty comes from the equation (\ref{final}).
Is there any simple way to compute this efficiently? 
Maybe it is helpful to use the
8-dimensional description for the internal space 
given in \cite{AI02} rather than
7-dimensional description.

Moreover, there should be  
$SO(7)^{-}$ invariant flow which contains the solution of 
\cite{Englert} and $SU(4)^{-}$ invariant flow which should contain the
solution of \cite{PW84} similarly.
Both of them are nonsupersymmetric. In principle, there will be no
problem for the former although it is rather invloved. 
However, for the latter, it is not known how
to construct the domain wall solution yet.

Eventually, one needs to understand 
the 11-dimensional lift of the whole $SU(3)$ invariant
flow which cover all of these supersymmetric or nonsupersymmetric flows. 
The present work will give some hints how to obtain the nontrivial
4-form field strengths. The main input is the 28-beins characterized
by four supergravity fields, the construction of Killing vectors and
the 11-dimensional metric. 
It would be an interesting open problem to find out this explicitly.

\vspace{.7cm}

\centerline{\bf Acknowledgments}

We would like to thank H. Nicolai
for correspondence and  T. Nishioka and B.R. Safdi on 
the 11-dimensional supersymmetry checking for discussions. 
This work was supported by Mid-career Researcher Program through
the National Research Foundation of Korea(NRF) grant 
funded by the Korea government(MEST)(No. 2009-0084601).
CA acknowledges warm hospitality and partial support of 
the Department of Physics, Princeton University(I.R. Klebanov).

\newpage

\appendix

\renewcommand{\thesection}{\large \bf \mbox{Appendix~}\Alph{section}}
\renewcommand{\theequation}{\Alph{section}\mbox{.}\arabic{equation}}

\section{The $SO(8)$ critical point }

The nonzero components where $m,n,p,q = 1, 2, \cdots, 7$
of the right hand side of (\ref{final}) for 
$SO(8)$ critical point can be summarized by
\bea
[12][12] & = & -\frac{12}{7} \, i \, \sqrt{2} \, s_{\theta}^4 \, s_{2}^2 \,
s_{3}^4 \, s_{4}^4 \, s_{5}^4 \, s_{6}^4 \, m_7, 
\qquad
[13][13]  =  -\frac{12}{7} \, i \, \sqrt{2} \, s_{\theta}^4 \, s_{2}^2 \,
s_{3}^2 \, s_{4}^4 \, s_{5}^4 \, s_{6}^4 \, m_7,
\nonu \\
{[}14][14] & = & -\frac{12}{7} \, i \, \sqrt{2} \, s_{\theta}^4 \, s_{2}^2 \,
s_{3}^2 \, s_{4}^2 \, s_{5}^4 \, s_{6}^4 \, m_7,
\qquad
[15][15]  =  -\frac{12}{7} \, i \, \sqrt{2} \, s_{\theta}^4 \, s_{2}^2 \,
s_{3}^2 \, s_{4}^2 \, s_{5}^2 \, s_{6}^4 \, m_7,
\nonu \\
{[}16][16] & = & -\frac{12}{7} \, i \, \sqrt{2} \, s_{\theta}^4 \, s_{2}^2 \,
s_{3}^2 \, s_{4}^2 \, s_{5}^2 \, s_{6}^2 \, m_7,
\qquad
[17][17]  =  -\frac{12}{7} \, i \, \sqrt{2} \, s_{\theta}^2 \, s_{2}^2 \,
s_{3}^2 \, s_{4}^2 \, s_{5}^2 \, s_{6}^2 \, m_7,
\nonu \\
{[}23][23] & = & -\frac{12}{7} \, i \, \sqrt{2} \, s_{\theta}^4 \,
s_{3}^2 \, s_{4}^4 \, s_{5}^4 \, s_{6}^4 \, m_7,
\qquad
[24][24]  =  -\frac{12}{7} \, i \, \sqrt{2} \, s_{\theta}^4 \,
s_{3}^2 \, s_{4}^2 \, s_{5}^4 \, s_{6}^4 \, m_7,
\nonu \\
{[}25][25] & = & -\frac{12}{7} \, i \, \sqrt{2} \, s_{\theta}^4 \,
s_{3}^2 \, s_{4}^2 \, s_{5}^2 \, s_{6}^4 \, m_7,
\qquad
[26][26]  =  -\frac{12}{7} \, i \, \sqrt{2} \, s_{\theta}^4 \,
s_{3}^2 \, s_{4}^2 \, s_{5}^2 \, s_{6}^2 \, m_7,
\nonu \\
{[}27][27] & = & -\frac{12}{7} \, i \, \sqrt{2} \, s_{\theta}^2 \,
s_{3}^2 \, s_{4}^2 \, s_{5}^2 \, s_{6}^2 \, m_7,
\qquad
[34][34]  =  -\frac{12}{7} \, i \, \sqrt{2} \, s_{\theta}^4 \,
 s_{4}^2 \, s_{5}^4 \, s_{6}^4 \, m_7,
\nonu \\
{[}35][35] & = & -\frac{12}{7} \, i \, \sqrt{2} \, s_{\theta}^4 \,
 s_{4}^2 \, s_{5}^2 \, s_{6}^4 \, m_7,
\qquad
[36][36]  =  -\frac{12}{7} \, i \, \sqrt{2} \, s_{\theta}^4 \,
 s_{4}^2 \, s_{5}^2 \, s_{6}^2 \, m_7,
\nonu \\
{[}37][37] & = & -\frac{12}{7} \, i \, \sqrt{2} \, s_{\theta}^2 \,
 s_{4}^2 \, s_{5}^2 \, s_{6}^2 \, m_7,
\qquad
[45][45]  =  -\frac{12}{7} \, i \, \sqrt{2} \, s_{\theta}^4 \,
s_{5}^2 \, s_{6}^4 \, m_7,
\nonu \\
{[}46][46] & = & -\frac{12}{7} \, i \, \sqrt{2} \, s_{\theta}^4 \,
s_{5}^2 \, s_{6}^2 \, m_7,
\qquad
[47][47]  =  -\frac{12}{7} \, i \, \sqrt{2} \, s_{\theta}^2 \,
s_{5}^2 \, s_{6}^2 \, m_7,
\nonu \\
{[}56][56] & = & -\frac{12}{7} \, i \, \sqrt{2} \, s_{\theta}^4 \,
s_{6}^2 \, m_7,
\qquad
[57][57]  =  -\frac{12}{7} \, i \, \sqrt{2} \, s_{\theta}^2 \,
s_{6}^2 \, m_7,
\nonu \\
{[}67][67] & = & -\frac{12}{7} \, i \, \sqrt{2} \, s_{\theta}^2  \, m_7.
\label{so8cri}
\eea
The antisymmetric notation for $[12][12]$ has  the components
of $1221$, $2112$ and $2121$ also and the first two are the same as $1212$ with
minus sign and the last one is the same as $1212$.

\section{The $SO(7)^{+}$ critical point }

The nonzero components of the right hand side of (\ref{final})
together with (\ref{newA}) for 
$SO(7)^{+}$ critical point can be summarized by
\bea
[12][12] & = & -\frac{2 i \sqrt{2}\cdot 5^{\frac{1}{4}}}{7}  
\, s_{\theta}^4 \, s_{2}^2 \,
s_{3}^4 \, s_{4}^4 \, s_{5}^4 \, s_{6}^4 \, m_7, 
\qquad
[13][13]  =  
 -\frac{2 i \sqrt{2}\cdot 5^{\frac{1}{4}} }{7}
\, s_{\theta}^4 \, s_{2}^2 \,
s_{3}^2 \, s_{4}^4 \, s_{5}^4 \, s_{6}^4 \, m_7,
\nonu \\
{[}14][14] & = & 
 -\frac{2 i \sqrt{2}\cdot 5^{\frac{1}{4}} }{7}
s_{\theta}^4 \, s_{2}^2 \,
s_{3}^2 \, s_{4}^2 \, s_{5}^4 \, s_{6}^4 \, m_7,
\qquad
[15][15]  =  
 -\frac{2 i \sqrt{2}\cdot 5^{\frac{1}{4}} }{7}
\, s_{\theta}^4 \, s_{2}^2 \,
s_{3}^2 \, s_{4}^2 \, s_{5}^2 \, s_{6}^4 \, m_7,
\nonu \\
{[}16][16] & = & 
 -\frac{2 i \sqrt{2}\cdot 5^{\frac{1}{4}} }{7}
\, s_{\theta}^4 \, s_{2}^2 \,
s_{3}^2 \, s_{4}^2 \, s_{5}^2 \, s_{6}^2 \, m_7,
\qquad
[17][17]  =  -\frac{2 i  \sqrt{2}
(3+2 c_{2\th})}{7 \cdot 5^{\frac{3}{4}}}   s_{\theta}^2  s_{2}^2 
s_{3}^2  s_{4}^2  s_{5}^2  s_{6}^2  m_7,
\nonu \\
{[}23][23] & = & 
 -\frac{2 i \sqrt{2}\cdot 5^{\frac{1}{4}} }{7}
\, s_{\theta}^4 \,
s_{3}^2 \, s_{4}^4 \, s_{5}^4 \, s_{6}^4 \, m_7,
\qquad
[24][24]  =  
 -\frac{2 i \sqrt{2}\cdot 5^{\frac{1}{4}} }{7}
s_{\theta}^4 \,
s_{3}^2 \, s_{4}^2 \, s_{5}^4 \, s_{6}^4 \, m_7,
\nonu \\
{[}25][25] & = & 
 -\frac{2 i \sqrt{2}\cdot 5^{\frac{1}{4}} }{7}
\, s_{\theta}^4 \,
s_{3}^2 \, s_{4}^2 \, s_{5}^2 \, s_{6}^4 \, m_7,
\qquad
[26][26]  =  
 -\frac{2 i \sqrt{2}\cdot 5^{\frac{1}{4}} }{7}
s_{\theta}^4 \,
s_{3}^2 \, s_{4}^2 \, s_{5}^2 \, s_{6}^2 \, m_7,
\nonu \\
{[}27][27] & = & 
 -\frac{2 i  \sqrt{2}
(3+2 c_{2\th})}{7\cdot 5^{\frac{3}{4}}} \,
 s_{\theta}^2 \,
s_{3}^2  \, s_{4}^2  \, s_{5}^2  \, s_{6}^2  \, m_7,
\qquad
[34][34]  =  
 -\frac{2 i \sqrt{2}\cdot 5^{\frac{1}{4}} }{7}
s_{\theta}^4 \,
 s_{4}^2 \, s_{5}^4 \, s_{6}^4 \, m_7,
\nonu \\
{[}35][35] & = & 
 -\frac{2 i \sqrt{2}\cdot 5^{\frac{1}{4}} }{7}
\, s_{\theta}^4 \,
 s_{4}^2 \, s_{5}^2 \, s_{6}^4 \, m_7,
\qquad
[36][36]  = 
 -\frac{2 i \sqrt{2}\cdot 5^{\frac{1}{4}} }{7}
\, s_{\theta}^4 \,
 s_{4}^2 \, s_{5}^2 \, s_{6}^2 \, m_7,
\nonu \\
{[}37][37] & = &  -\frac{2 i  \sqrt{2}
(3+2 c_{2\th})}{7\cdot 5^{\frac{3}{4}}} \,
s_{\theta}^2 \,
 s_{4}^2  \, s_{5}^2  \, s_{6}^2  \, m_7,
\qquad
[45][45]  = 
 -\frac{2 i \sqrt{2}\cdot 5^{\frac{1}{4}} }{7}
\, s_{\theta}^4 \,
s_{5}^2 \, s_{6}^4 \, m_7,
\nonu \\
{[}46][46] & = & 
 -\frac{2 i \sqrt{2}\cdot 5^{\frac{1}{4}} }{7}
\, s_{\theta}^4 \,
s_{5}^2 \, s_{6}^2 \, m_7,
\qquad
[47][47]  =   -\frac{2 i  \sqrt{2}
(3+2c_{ 2\th})}{7\cdot 5^{\frac{3}{4}}}
\, s_{\theta}^2 \,
s_{5}^2  \, s_{6}^2  \, m_7,
\nonu \\
{[}56][56] & = & 
 -\frac{2 i \sqrt{2}\cdot 5^{\frac{1}{4}} }{7}
\, s_{\theta}^4 \,
s_{6}^2 \, m_7,
\qquad
[57][57]  =  -\frac{2 i  \sqrt{2}
(3+2c_{ 2\th})}{7\cdot 5^{\frac{3}{4}}} 
\, s_{\theta}^2 \, 
s_{6}^2  \, m_7,
\nonu \\
{[}67][67] & = & -\frac{2 i  \sqrt{2}
(3+2c_{ 2\th})}{7\cdot 5^{\frac{3}{4}}}
\, s_{\theta}^2  \, m_7.
\label{so7cri}
\eea
The expressions having the index $7$ contain the quantity
$(3+2 \cos 2\th)$ which is proportional to $\xi^2$ from (\ref{xi}) and
(\ref{warp1}).
Furthermore, the metric is given by (\ref{7dmetric}) where the $(7,7)$
component of the metric has $\xi^2$ dependence.

\section{The generalized vielbeins}

Let us present the generalized vielbeins $e^m_{ij}=e^{mij}$ as follows.
For $m=1$, there are
\bea
e^{1}_{12} & = & -\frac{2}{\sqrt{a(r)}} s^{-1}_{2} s^{-1}_{3} \left[
\frac{c_{4}}{s_{4}} s_{1}+ c_{1}  \frac{c_{5}}{s_{5}}  s^{-1}_{4} \right],
\nonu \\
e^{1}_{13} & = & \sqrt{\frac{2}{a(r)}} \left[s_1
s^{-1}_{2}(\frac{c_{3}}{s_{3}}+\frac{c_{6}}{s_{6}}s^{-1}_{3}
 s^{-1}_{4}s^{-1}_{5})+c_{1}(-\frac{c_{2}}{s_{2}}+a(r)^{2}\frac{c_{\theta}}{s_{\theta}}
 s^{-1}_{2} s^{-1}_{3}s^{-1}_{4}s^{-1}_{5}s^{-1}_{6}) \right] ,
\nonu \\
e^{1}_{14} & = &-\sqrt{\frac{2}{a(r)}}\left[ \frac{c_{2}}{s_{2}}
s_1+s_2^{-1} (c_1 ( \frac{c_{3}}{s_{3}}+ \frac{c_{6}}{s_{6}}
s_3^{-1} s_4^{-1} s_5^{-1})-a(r)^2 \frac{c_{\theta}}{s_{\theta}} s_1
s_3^{-1} s_4^{-1} s_5^{-1} s_6^{-1}) \right],
\nonu \\
e^{1}_{15} & = &\frac{2}{\sqrt{a(r)}} s^{-1}_{2} s^{-1}_{3} \left[
c_{1}\frac{c_{4}}{s_{4}} -  \frac{c_{5}}{s_{5}}s_{1}   s^{-1}_{4} \right],
\qquad
 e^{1}_{16}  = \frac{2}{\sqrt{a(r)}} ,
\nonu \\
e^{1}_{17} & = &-\sqrt{\frac{2}{a(r)}} \left[s_1
s^{-1}_{2}(\frac{c_{3}}{s_{3}}-\frac{c_{6}}{s_{6}}s^{-1}_{3}
 s^{-1}_{4}s^{-1}_{5})+c_{1}(\frac{c_{2}}{s_{2}}+a(r)^{2}\frac{c_{\theta}}{s_{\theta}}
 s^{-1}_{2} s^{-1}_{3}s^{-1}_{4}s^{-1}_{5}s^{-1}_{6}) \right]  ,
\nonu \\
e^{1}_{18} & = &\sqrt{\frac{2}{a(r)}}\left[ \frac{c_{2}}{s_{2}} s_1+s_2^{-1}
(c_1 (- \frac{c_{3}}{s_{3}}+ \frac{c_{6}}{s_{6}} s_3^{-1} s_4^{-1}
s_5^{-1})+a(r)^2 \frac{c_{\theta}}{s_{\theta}} s_1 s_3^{-1} s_4^{-1}
s_5^{-1} s_6^{-1}) \right],
\nonu \\
e^{1}_{23} & = &\sqrt{\frac{2}{a(r)}} \left[ \frac{c_{2}}{s_{2}} s_1+s_2^{-1}
(c_1 ( \frac{c_{3}}{s_{3}}- \frac{c_{6}}{s_{6}} s_3^{-1} s_4^{-1}
s_5^{-1})+a(r)^2 \frac{c_{\theta}}{s_{\theta}} s_1 s_3^{-1} s_4^{-1}
s_5^{-1} s_6^{-1}) \right],
\nonu \\
e^{1}_{24} & = & \sqrt{\frac{2}{a(r)}} \left[s_1
s^{-1}_{2}(\frac{c_{3}}{s_{3}}-\frac{c_{6}}{s_{6}}s^{-1}_{3}
 s^{-1}_{4}s^{-1}_{5})-c_{1}(\frac{c_{2}}{s_{2}}+a(r)^{2}\frac{c_{\theta}}{s_{\theta}}
 s^{-1}_{2} s^{-1}_{3}s^{-1}_{4}s^{-1}_{5}s^{-1}_{6}) \right] ,
\nonu \\
e^{1}_{25} & = &-e^1_{16},
\qquad
 e^{1}_{26}  = e^1_{15},
\nonu \\
e^{1}_{27} & = &\sqrt{\frac{2}{a(r)}} \left[ -\frac{c_{2}}{s_{2}}
s_1+s_2^{-1} (c_1 ( \frac{c_{3}}{s_{3}}+ \frac{c_{6}}{s_{6}}
s_3^{-1} s_4^{-1} s_5^{-1})+a(r)^2 \frac{c_{\theta}}{s_{\theta}} s_1
s_3^{-1} s_4^{-1} s_5^{-1} s_6^{-1}) \right],
\nonu \\
e^{1}_{28} & = & -\sqrt{\frac{2}{a(r)}} \left[s_1
s^{-1}_{2}(\frac{c_{3}}{s_{3}}+\frac{c_{6}}{s_{6}}s^{-1}_{3}
 s^{-1}_{4}s^{-1}_{5})+c_{1}(\frac{c_{2}}{s_{2}}-a(r)^{2}\frac{c_{\theta}}{s_{\theta}}
 s^{-1}_{2} s^{-1}_{3}s^{-1}_{4}s^{-1}_{5}s^{-1}_{6})\right] ,
\nonu \\
e^{1}_{34} & = &- e^1_{16},
\qquad
 e^{1}_{35}  = -e^1_{24},
\qquad
e^{1}_{36}  =  e^1_{14},
\nonu \\
e^{1}_{37} & = & \frac{2}{\sqrt{a(r)}} s^{-1}_{2} s^{-1}_{3} \left[
c_{1}\frac{c_{4}}{s_{4}}+ \frac{c_{5}}{s_{5}} s_{1} s^{-1}_{4} \right],
\qquad
 e^{1}_{38}  =  \frac{2}{\sqrt{a(r)}} s^{-1}_{2} s^{-1}_{3} \left[-
\frac{c_{4}}{s_{4}}s_{1}+ c_{1}\frac{c_{5}}{s_{5}}  s^{-1}_{4} \right],
\nonu \\
e^{1}_{45} & = & e^1_{23},
\qquad
e^{1}_{46}  =   -e^1_{13},
\qquad
e^{1}_{47}  =   -e^1_{38},
\qquad
 e^{1}_{48}  =  e^1_{37},
\qquad
e^{1}_{56}  =  -e^1_{12},
\nonu \\
e^{1}_{57}  & = &  -e^1_{28},
\qquad
e^{1}_{58}  =  e^1_{27},
\qquad
e^{1}_{67}  =  e^1_{18},
\qquad
e^{1}_{68}  =  -e^1_{17},
\qquad
e^{1}_{78}  =  e^{1}_{16}.
\nonu
\eea
For other values for $m$,
the generalized vielbeins we do not present(for simplicity) here can be
constructed from (\ref{w}), (\ref{K}), (\ref{rect}), and (\ref{emij}). 
The indices $i,j$ in the generalized vielbein are antisymmetric. 

\section{The $SO(7)^{+}$ invariant flow }

Moreover, the nonzero quantities of 
the right hand side of (\ref{final1}) for fixed $\mu=r$ are given by
\bea
{[}117] & = & -\frac{1024 \, i  \, a(r)^{\frac{1}{2}} \, \left[ a(r)^4 -
1\right]}{[1+a(r)^4 +(-1+a(r)^4)c_{2\th}]} \, c_{\theta} \,  s_{\theta}^3 \, s_{2}^2 \,
s_{3}^2 \, s_{4}^2 \, s_{5}^2 \, s_{6}^2 \, m_7, \nonu \\
{[}227] & = & 
-\frac{1024 \, i  \, a(r)^{\frac{1}{2}} \, \left[ a(r)^4 -
1\right]}{[1+a(r)^4 +(-1+a(r)^4)c_{ 2\th}]}
\, c_{\theta} \,  s_{\theta}^3 \, 
s_{3}^2 \, s_{4}^2 \, s_{5}^2 \, s_{6}^2 \, m_7, \nonu \\
{[}337] & = & 
-\frac{1024 \, i  \, a(r)^{\frac{1}{2}} \, \left[ a(r)^4 -
1\right]}{[1+a(r)^4 +(-1+a(r)^4)c_{ 2\th}]}
\, c_{\theta} \,  s_{\theta}^3 
\, s_{4}^2 \, s_{5}^2 \, s_{6}^2 \, m_7, 
\nonu \\
{[}447] & = & 
-\frac{1024 \, i  \, a(r)^{\frac{1}{2}} \, \left[ a(r)^4 -
1\right]}{[1+a(r)^4 +(-1+a(r)^4)c_{ 2\th}]}
\, c_{\theta} \,  s_{\theta}^3 
\, s_{5}^2 \, s_{6}^2, \, m_7, \nonu \\
{[}557] & = & 
-\frac{1024 \, i  \, a(r)^{\frac{1}{2}} \, \left[ a(r)^4 -
1\right]}{[1+a(r)^4 +(-1+a(r)^4) c_{2\th}]}
\, c_{\theta} \,  s_{\theta}^3
\, s_{6}^2,\, m_7, 
\nonu \\
{[}667] & = & 
-\frac{1024 \, i  \, a(r)^{\frac{1}{2}} \, \left[ a(r)^4 -
1\right]}{[1+a(r)^4 +(-1+a(r)^4) c_{2\th}]}
\, c_{\theta} \,  s_{\theta}^3 \, m_7.
\label{nonzero}
\eea
Note that the overall factor $\sin \theta \cos \theta$ appears in
these expressions and this plays the  role of $\theta$-dependence in
the 4-form (\ref{12311}).

The nonzero components of the right hand side of (\ref{final}) where
the equation (\ref{newA}) is used for 
$SO(7)^{+}$ invariant flow can be summarized by
\bea
[12][12] & = & -\frac{2 i \sqrt{2}\cdot a(r)}{7}  
\, s_{\theta}^4 \, s_{2}^2 \,
s_{3}^4 \, s_{4}^4 \, s_{5}^4 \, s_{6}^4 \, m_7, 
\qquad
[13][13]  =  
 -\frac{2 i \sqrt{2}\cdot a(r) }{7}
\, s_{\theta}^4 \, s_{2}^2 \,
s_{3}^2 \, s_{4}^4 \, s_{5}^4 \, s_{6}^4 \, m_7,
\nonu \\
{[}14][14] & = & 
 -\frac{2 i \sqrt{2}\cdot a(r) }{7}
s_{\theta}^4 \, s_{2}^2 \,
s_{3}^2 \, s_{4}^2 \, s_{5}^4 \, s_{6}^4 \, m_7,
\qquad
[15][15]  =  
 -\frac{2 i \sqrt{2}\cdot a(r) }{7}
\, s_{\theta}^4 \, s_{2}^2 \,
s_{3}^2 \, s_{4}^2 \, s_{5}^2 \, s_{6}^4 \, m_7,
\nonu \\
{[}16][16] & = & 
 -\frac{2 i \sqrt{2}\cdot a(r) }{7}
\, s_{\theta}^4 \, s_{2}^2 \,
s_{3}^2 \, s_{4}^2 \, s_{5}^2 \, s_{6}^2 \, m_7,
\nonu \\
{[}17][17]  & = &  -\frac{2 i  \sqrt{2}
[1+a(r)^4 +(-1+a(r)^4)c_{ 2\th}]}{7 \cdot a(r)^{3}}   s_{\theta}^2  s_{2}^2 
s_{3}^2  s_{4}^2  s_{5}^2  s_{6}^2  m_7,
\nonu \\
{[}23][23] & = & 
 -\frac{2 i \sqrt{2}\cdot a(r) }{7}
\, s_{\theta}^4 \,
s_{3}^2 \, s_{4}^4 \, s_{5}^4 \, s_{6}^4 \, m_7,
\qquad
[24][24]  =  
 -\frac{2 i \sqrt{2}\cdot a(r) }{7}
s_{\theta}^4 \,
s_{3}^2 \, s_{4}^2 \, s_{5}^4 \, s_{6}^4 \, m_7,
\nonu \\
{[}25][25] & = & 
 -\frac{2 i \sqrt{2}\cdot a(r) }{7}
\, s_{\theta}^4 \,
s_{3}^2 \, s_{4}^2 \, s_{5}^2 \, s_{6}^4 \, m_7,
\qquad
[26][26]  =  
 -\frac{2 i \sqrt{2}\cdot a(r) }{7}
s_{\theta}^4 \,
s_{3}^2 \, s_{4}^2 \, s_{5}^2 \, s_{6}^2 \, m_7,
\nonu \\
{[}27][27] & = & 
 -\frac{2 i  \sqrt{2}
[1+a(r)^4 +(-1+a(r)^4)c_{ 2\th}]}{7\cdot a(r)^{3}} \,
 s_{\theta}^2 \,
s_{3}^2  \, s_{4}^2  \, s_{5}^2  \, s_{6}^2  \, m_7,
\nonu \\
{[}34][34]  & = &  
 -\frac{2 i \sqrt{2}\cdot a(r) }{7}
s_{\theta}^4 \,
 s_{4}^2 \, s_{5}^4 \, s_{6}^4 \, m_7,
\nonu \\
{[}35][35] & = & 
 -\frac{2 i \sqrt{2}\cdot a(r) }{7}
\, s_{\theta}^4 \,
 s_{4}^2 \, s_{5}^2 \, s_{6}^4 \, m_7,
\qquad
[36][36]  = 
 -\frac{2 i \sqrt{2}\cdot a(r) }{7}
\, s_{\theta}^4 \,
 s_{4}^2 \, s_{5}^2 \, s_{6}^2 \, m_7,
\nonu \\
{[}37][37] & = &  -\frac{2 i  \sqrt{2}
[1+a(r)^4 +(-1+a(r)^4)c_{ 2\th}]}{7\cdot a(r)^{3}} \,
s_{\theta}^2 \,
 s_{4}^2  \, s_{5}^2  \, s_{6}^2  \, m_7,
\nonu \\
{[}45][45]  & = & 
 -\frac{2 i \sqrt{2}\cdot a(r)}{7}
\, s_{\theta}^4 \,
s_{5}^2 \, s_{6}^4 \, m_7,
\qquad
{[}46][46]  =  
 -\frac{2 i \sqrt{2}\cdot a(r) }{7}
\, s_{\theta}^4 \,
s_{5}^2 \, s_{6}^2 \, m_7,
\nonu \\
{[}47][47]  & = &   -\frac{2 i  \sqrt{2}
[1+a(r)^4 +(-1+a(r)^4)c_{ 2\th}]}{7\cdot a(r)^{3}}
\, s_{\theta}^2 \,
s_{5}^2  \, s_{6}^2  \, m_7,
\nonu \\
{[}56][56] & = & 
 -\frac{2 i \sqrt{2}\cdot a(r) }{7}
\, s_{\theta}^4 \,
s_{6}^2 \, m_7,
\qquad
[57][57]  =  -\frac{2 i  \sqrt{2}
[1+a(r)^4 +(-1+a(r)^4)c_{ 2\th}]}{7\cdot a(r)^{3}} 
\, s_{\theta}^2 \, 
s_{6}^2  \, m_7,
\nonu \\
{[}67][67] & = & -\frac{2 i  \sqrt{2}
[1+a(r)^4 +(-1+a(r)^4)c_{ 2\th}]}{7\cdot a(r)^{3}}
\, s_{\theta}^2  \, m_7.
\label{so7flow}
\eea
Of course, at the critical point condition (\ref{vev}),
the above results (\ref{so7flow}) reduces to the ones in (\ref{so7cri}).
For the $SO(8)$ critical point($a(r)=1$), this (\ref{vev}) becomes 
the ones in (\ref{so8cri}) except the overall factor. 

\section{The supersymmetry transformation for 11-dimensional solutions}

The supersymmetry transformation rule \cite{DNP} of the gravitino of
11-dimensional supergravity becomes 
in a purely bosonic background 
\bea
\delta \Psi_M = {\cal D}_M \, \epsilon,
\nonu
\eea
where $\epsilon$ is an anticommuting parameter and 
\bea
{\cal D}_M = D_M -\frac{i}{144} \left( \Gamma_M^{\,\,\,NPQR} -8
  \delta_M^N \, \Gamma^{PQR} \right) \, F_{NPQR}, \qquad
D_M = \partial_M -\frac{1}{4} \omega_M^{\,\,\, AB} \, \Gamma_{AB}.
\label{der}
\eea
Sometimes the numerical factor in front of the 4-forms  in (\ref{der})
is different due to the different normalization of 4-forms in the
11-dimensional Einstein-Maxwell equations.
The number of supersymmetries preserved by an 11-dimensional
background
depends on the number of covariantly constant spinors 
${\cal D}_M \, \epsilon=0$ called Killing spinors. The Killing spinors 
satisfy the integrability condition
\bea
[{\cal D}_M, {\cal D}_N] \, \epsilon =0.
\label{int}
\eea
This is a necessary but not sufficient condition for the existence of
Killing spinors.

Let us look at this condition (\ref{int}) closely. 
The condition for zero torsion leads to the fact that 
the spin connection is given by
\bea
\omega_{MAB} = \frac{1}{2} \left( -\Omega_{MAB} + \Omega_{ABM}
  -\Omega_{BMA} 
\right),
\label{conn}
\eea
where $
\Omega_{MN}^{\,\,\,\,\,\,\,\,\,A} = - 2 \pa_{[M} \, e_{N]}^{A}$.
The first term of (\ref{conn}) can be obtained 
from the vielbein, the $\Omega_{MN}^{\,\,\,\,\,\,\,\,\,A}$  and 
the metric $\eta_{AB} =\mbox{diag}(-1, 1, \cdots, 1)= \eta^{AB}$
and the last two terms of (\ref{conn}) can be written
in terms of the first term as follows:
\bea
\Omega_{MAB} = e_{A}^{N} \, \Omega_{MN}^{\,\,\,\,\,\,\,\,\,C} \,
\eta_{CB}, \qquad
 \Omega_{ABM} =
 e_{A}^{N} \, \Omega_{NBC} \,
e_{M}^{C}, \qquad
 \Omega_{BMA} =
 e_{B}^{N} \, \Omega_{NCA} \, e_{M}^{C}.
\label{Ome3}
\eea
Then from (\ref{Ome3}), one gets the final expression for (\ref{conn}).
In (\ref{der}), the spin connection can be written in terms of (\ref{conn})
$
\omega_{M}^{\,\,\,AB} = \omega_{MCD} \, \eta^{CA} \, \eta^{DB}$.
We want to write (\ref{der}) in the frame basis(tangent space indices) 
for simplicity of (\ref{der})(Gamma matrix with world indices are
complicated but those with tangent indices are constant) and we need to have
the relation 
$
\omega_{A}^{\,\,\,BC} = e_A^{M} \, \omega_M^{\,\,\,BC}$.
Let us decompose (\ref{der}) into the $\pa_M$ term and the other. By
multiplying the vielbein $e_{A}^{M}$, the latter can be
written as  
\bea
\widetilde{\omega}_A \equiv   -\frac{1}{4} \omega_A^{\,\,\, BC} \,
\Gamma_{BC}
-\frac{i}{144} \left( \Gamma_A^{\,\,\,BCDE} -8
  \delta_A^B \, \Gamma^{CDE} \right) \, F_{BCDE},
\label{fun}
\eea
where as usual, the 4-form in frame basis 
is related to the one in coordinate basis:$F_{ABCD}=
e_{A}^{M} \, e_{B}^{N} \, e_{C}^{P}\, e_{D}^{Q}\, F_{MNPQ}$. The Gamma
matrix is given in \cite{AW01,AI}.
By multiplying $e_A^{M} \, e_B^{N}$ into (\ref{int}), 
one gets, by carefully reorganizing it, 
the following integrability condition(Note 11-dimensional spacetime
coordinates are decomposed into as follows:$z^M=(x^{\mu}, y^m)$) 
\bea
\left(
e_{A}^{M} \, \frac{\pa \widetilde{\omega}_B}{\pa z^M}
-e_A^{M} \, 
\frac{\pa e_B^{N}}{\pa z^M} \, e_N^{C} \, \widetilde{\omega}_C
-e_{B}^{N} \, \frac{\pa \widetilde{\omega}_A}{\pa z^N}
+e_B^{M} \, \frac{\pa e_A^{N}}{\pa 
z^M} \, e_N^{C} \, \widetilde{\omega}_C +[\widetilde{\omega}_A,
\widetilde{\omega}_B]
\right) \, \epsilon =0.
\label{newint}
\eea

Now we are ready to compute (\ref{newint}) with (\ref{fun}) and
vielbein which can be obtained from 11-dimensional metric (\ref{11d}) and (\ref{7dmetric}).
For example, 
one can compute the $32 \times 32$ matrix elements of $[{\cal D}_1,
{\cal D}_5]$ in (\ref{newint}). The nonzero expressions of them are summarized
by
the following matrix elements
$(1,2)$, $(1,18)$, $(3,4)$, $(3,20)$, $(6,5)$, $(6,21)$, $(8,7)$, 
$(8,23)$, $(9,26)$, $(10,9)$,
$(11,28)$, $(12,11)$,
$(13,14)$, $(14,29)$, $(15,16)$, $(16,31)$, $(18,1)$, $(18,17)$,
$(20,3)$,
$(20,19)$, $(21,6)$, $(21,22)$,
$(23,8)$, $(23,24)$, $(25,26)$, $(26,9)$, $(27,28)$, $(28,11)$,
$(29,14)$,
$(30,29)$, $(31,16)$, $(32,31)$,
which have the nonzero value
\bea
-\frac{[a(r)^4-1][a(r)^4+5]}{36 \, L^2 \, a(r)^{\frac{2}{3}}\,
[a(r)^4 \, c_\theta^2 +s_\theta^2]^{\frac{2}{3}}}.
\label{value}
\eea
The transpose elements have same nonzero values except minus sign.
After substituting (\ref{value}) into (\ref{newint}) for 32 components
of $\epsilon$, the half of them are fixed. Now we move on the 
matrix elements of $[{\cal D}_4,
{\cal D}_5]$ in (\ref{newint}). 
The nonzero expressions of them are summarized
by
the following matrix elements
$(1,2)$, $(2,17)$, $(3,4)$, $(4,19)$, $(5,22)$, $(6,5)$, $(7,24)$,
$(8,7)$, $(9,10)$, $(9,26)$,
$(11,12)$, $(11,28)$,
$(14,13)$, $(14,29)$, $(16,15)$, $(16,31)$, $(17,18)$,
$(18,1)$, $(19,20)$,
$(20,3)$,
$(21,6)$, $(22,21)$,
$(23,8)$, $(24,23)$, $(25,10)$, $(25,26)$, $(27,12)$,
$(27,28)$, $(30,13)$, $(30,29)$, $(32,15)$,
$(32,31)$,
which have the nonzero value
\bea
-\frac{[a(r)^4-1]\, a(r)^{\frac{10}{3}}\, [6 a(r)^8 \, c_\theta^4 - 2
  a(r)^4 \, c_\theta^2 (-16 + c_{2\theta})+(-11+c_{2\theta} 
\, s_\theta^2 )]}{72 \, L^2 \,
[a(r)^4 \, c_\theta^2 +s_\theta^2]^{\frac{8}{3}}}.
\label{value1}
\eea
Furthermore, the following nonzero matrix elements
$(1,23)$, $(2,24)$, $(3,21)$, $(4,22)$, $(5,3)$, $(6,4)$,
$(7,1)$, $(8,2)$, $(13,11)$, $(13,27)$,
$(14,12)$, $(14,28)$,
$(15,9)$, $(15,25)$, $(16,10)$, $(16,26)$, $(17,7)$,
$(18,8)$, $(19,5)$, $(20,6)$,
$(21,19)$, $(22,20)$,
$(23,17)$, $(24,18)$, $(29,11)$, $(29,27)$,
$(30,12)$, $(30,28)$, $(31,9)$, $(31,25)$, $(32,10)$, $(32,26)$,
have the value
\bea
-\frac{[a(r)^4-1]\, a(r)^{\frac{4}{3}}\, [ a(r)^8 \, c_\theta^2 - 
  a(r)^4 \,(22 + 3 c_{2\theta})-5 s_\theta^2 )] \, s_{2\theta}}
{72 \, L^2 \,
[a(r)^4 \, c_\theta^2 +s_\theta^2]^{\frac{8}{3}}}.
\label{value2}
\eea

After substituting (\ref{value1})  and (\ref{value2}) into (\ref{newint}) for 32 components
of $\epsilon$, the remaining  independent parameters are
vanishing. Therefore the supersymmetry is completely broken.
That is, along the RG flow, there is no supersymmetry except the 
$SO(8)$ critical point.  
As expected, at the $SO(8)$ critical point($a(r)=1$), one sees that 
the matrix elements (\ref{value}), (\ref{value1}) and (\ref{value2}) 
vanish and moreover, all the other matrix elements we do not present
here vanish and therefore, there exists  the maximal supersymmetries.
At the $SO(7)^{+}$ critical point, one can go back the equation
(\ref{sol})
and check the above supersymmetry condition and this gives the
nonsupersymmetric case. 


\end{document}